\documentclass[aps,prx,amsmath,amssymb,reprint,superscriptaddress]{revtex4-2}
\usepackage[utf8]{inputenc}
\usepackage[T1]{fontenc}
\usepackage{lmodern}
\usepackage{graphicx}% Include figure files
\usepackage[colorlinks,linkcolor=blue]{hyperref}
\usepackage[nameinlink,capitalise]{cleveref}
\usepackage{braket}
\usepackage{dsfont}

\begin{document}

\title{Identification of different types of high-frequency defects in superconducting qubits} %Title of paper

\author{Leonid V. Abdurakhimov}
\email{leonid.abdurakhimov@meetiqm.com}
\altaffiliation[present address: ]{IQM Finland Oy, Espoo 02150, Finland}
\affiliation{NTT Basic Research Laboratories, NTT Corporation, 3-1 Morinosato-Wakamiya, Atsugi, Kanagawa 243-0198, Japan}

\author{Imran Mahboob}
\affiliation{NTT Basic Research Laboratories, NTT Corporation, 3-1 Morinosato-Wakamiya, Atsugi, Kanagawa 243-0198, Japan}

\author{Hiraku Toida}
\affiliation{NTT Basic Research Laboratories, NTT Corporation, 3-1 Morinosato-Wakamiya, Atsugi, Kanagawa 243-0198, Japan}

\author{Kosuke Kakuyanagi}
\affiliation{NTT Basic Research Laboratories, NTT Corporation, 3-1 Morinosato-Wakamiya, Atsugi, Kanagawa 243-0198, Japan}

\author{Yuichiro Matsuzaki}
\affiliation{Research Center for Emerging Computing Technologies, National Institute of Advanced Industrial Science and Technology (AIST), 1-1-1 Umezono, Tsukuba, Ibaraki 305-8568, Japan}

\author{Shiro Saito}
\email{shiro.saito.bx@hco.ntt.co.jp}
\affiliation{NTT Basic Research Laboratories, NTT Corporation, 3-1 Morinosato-Wakamiya, Atsugi, Kanagawa 243-0198, Japan}

\date{\today}

\begin{abstract}
Parasitic two-level-system (TLS) defects are one of the major factors limiting the coherence times of superconducting qubits.
Although there has been significant progress in characterizing basic parameters of TLS defects, exact mechanisms of interactions between a qubit and various types of TLS defects remained largely unexplored due to the lack of experimental techniques able to probe the form of qubit-defect couplings.
Here we present an experimental method of TLS defect spectroscopy using a strong qubit drive that allowed us to distinguish between various types of qubit-defect interactions. 
By applying this method to a capacitively shunted flux qubit, we detected a rare type of TLS defect with a nonlinear qubit-defect coupling due to critical-current fluctuations, as well as conventional TLS defects with a linear coupling to the qubit caused by charge fluctuations. 
The presented approach could become the routine method for high-frequency defect inspection and quality control in superconducting qubit fabrication, providing essential feedback for fabrication process optimization.
The reported method is a powerful tool to uniquely identify the type of noise fluctuations caused by TLS defects, enabling the development of realistic noise models relevant to noisy intermediate-scale quantum (NISQ) computing and fault-tolerant quantum control. 
\end{abstract}

\maketitle %\maketitle must follow title, authors, abstract and \pacs

\section{Introduction}

Reducing gate error rates of a physical qubit is one of the biggest challenges on the road to building a gate-based fault-tolerant superconducting quantum computer~\cite{Kjaergaard2020}. For the typical physical error rate of $10^{-3}$, a logical qubit would need to contain about $10^3$ -- $10^4$ physical qubits in order to achieve sufficiently low logical error rates~\cite{Fowler2012}, and a full-scale realization of quantum applications would require more than 10$^7$ physical qubits using the standard surface-code approach~\cite{Gidney2021} (this requirement can be slightly relaxed by using hybrid architectures~\cite{Gouzien2021}). Potentially, useful tasks can be performed using a noisy intermediate-scale quantum (NISQ) processors consisting of tens to hundreds physical qubits~\cite{Preskill2018,Bharti2022}, but, even in this case, quantum error mitigation is crucial~\cite{Cao2021,Endo2021}. The value of a gate error rate can be estimated as the ratio of a gate duration time and a qubit coherence time, and significant efforts have been made to improve gate speeds and qubit coherence. Two-qubit gate error rates on the order of $10^{-3}$ were demonstrated for transmon qubits with coherence times exceeding 100 $\mu$s~\cite{Kandala2021}. The possibilities of the further improvement of coherence times were demonstrated using novel qubit designs, such as a fluxonium qubit~\cite{Zhang2021universal,somoroff2021millisecond}, a 0 -- $\pi$ qubit~\cite{Gyenis2021}, and bosonic qubits~\cite{Ofek2016,Grimm2020}. However, independent of a particular qubit architecture, coherence times are ultimately limited by intrinsic losses in materials constituting a qubit, and reducing the number of material defects is a crucial factor for the qubit performance improvement~\cite{Oliver2013,Place2021,Leon2021}.

Extensive studies in recent decades have identified major noise sources in superconducting qubits, including parasitic two-level-system (TLS) defects~\cite{Simmonds2004,Martinis2005,Martin2005,Lupascu2009,Grabovskij2012,Barends2013,Wang2015a,Klimov2018,Burnett2019,Muller2019,Lisenfeld2016, Lisenfeld2019,Bilmes2020,Bilmes2021,Andersson2021,bilmes2022}, quasiparticle noise~\cite{Yamamoto2006,Gustavsson2016,Hosseinkhani2017,Riwar2019,Vepsaelaeinen2020,Cardani2021,rafferty2021spurious,martinis2021,Wilen2021,mcewen2022,mannila2022}, and, for flux-tunable qubits, a $1/f$ magnetic flux noise~\cite{Wang2015,Kumar2016}. In this paper, we focus on TLS defects: defects of different microscopic nature that can be modeled as two-level quantum systems, such as tunneling atoms, dangling electronic bonds, impurity atoms, and trapped charge states~\cite{Muller2019}. TLS defects are ubiquitous in the current generation of superconducting qubits: in the most commonly used method of Josephson junction fabrication, a junction tunnel barrier is formed of an amorphous layer of aluminum oxide that hosts a large number of TLS defects~\cite{Muller2019,bilmes2022}. Besides, the microwave absorption due to TLS defects formed in various oxide layers is one of the major mechanisms limiting Q-factors of microwave superconducting resonators~\cite{Goetz2016,Altoe2020}, including superconducting 3D cavities which are used for quantum memory applications and bosonic quantum computing~\cite{Kudra2020,Romanenko2020,heidler2021observing}.
Normally, it is assumed that the qubit-defect interaction is caused by an electric-field coupling between an electric dipole associated with a charge TLS defect and a microwave electric field generated across a Josephson junction~\cite{Martinis2005}.

The standard techniques to study high-frequency TLS defects in superconducting qubits are based on matching the frequencies of qubit and defect transitions in the laboratory frame either by adjusting the qubit frequency with an applied magnetic flux~\cite{Simmonds2004,Martinis2005,Lupascu2009,Lisenfeld2010,Gustavsson2012,Barends2013}, or changing the defect frequency with an external electrical field or an applied mechanical strain~\cite{Grabovskij2012,Lisenfeld2019,Bilmes2020,Bilmes2021}. Parameters of a TLS defect are then extracted from the splitting of an avoided crossing observed in the qubit spectrum~\cite{Simmonds2004,Martinis2005}, the qubit response to a Rabi drive~\cite{Ashhab2006,Lisenfeld2010}, or measurements of energy-relaxation times $T_1$~\cite{Barends2013,Klimov2018,Burnett2019,Lisenfeld2019,Bilmes2020,Bilmes2021}. The standard methods of TLS defect spectroscopy based on the qubit frequency tuning are not suitable for fixed-frequency qubits, such as fixed-frequency transmons. Recently, it was shown that TLS defects in fixed-frequency transmons could be probed by utilizing an AC Stark shift of the qubit frequency induced by an \textit{off-resonant} microwave drive~\cite{Carroll2022}. However, the frequency band of the demonstrated technique was quite narrow ($\pm$25\,MHz). Importantly, previously described methods allowed to study basic parameters of TLS defects, but the exact mechanisms of qubit-defect interactions remained largely unexplored so far.

In this paper, we describe a method for the identification of different types of high-frequency TLS defects using a strong \textit{resonant} qubit drive. 
Our approach is based on the notion that the relaxation of a strongly driven qubit can be affected by the presence of an off-resonant high-frequency TLS defect due to the dynamic splitting of the qubit transition frequency, caused by the \textit{resonant} AC Stark effect (Autler-Townes splitting), and the corresponding dynamic matching of qubit and TLS defect frequencies in the rotating frame~\cite{Abdurakhimov2020}.
In the case of a flux-tunable qubit, the reported technique allows one to extract detailed information about the qubit-defect interaction, including the type of the qubit-defect coupling. 
We demonstrate the capabilities of the method by performing TLS defect spectroscopy in a capacitively shunted (c-shunt) flux qubit in a wide frequency band $\pm$120\,MHz.
By measuring the strongly-driven qubit state relaxation at different values of an external magnetic flux bias, we identified two different types of defects: standard charge-fluctuation TLS defects and TLS defects with a nonlinear coupling to the qubit mediated by critical-current fluctuations.

We show that, near the optimal magnetic flux bias point, the coupling between the qubit and a charge-fluctuation TLS defect has the form of a beam splitter type interaction, while a critical-current-fluctuation TLS defect is coupled to the qubit via a three-wave-mixing interaction term. Due to the different shapes of qubit-defect interactions, the dependence of the position of a charge-fluctuation TLS signature on the applied magnetic flux is determined by the corresponding shift of the qubit frequency, while the signature of a critical-current TLS defect follows the double value of the qubit frequency shift. This behavior forms the basis of the reported differentiation technique. In addition, by repeating the measurements with an applied in-plane magnetic field, we can exclude the possibility that observed TLS signatures are caused by magnetic TLS defects.

The reported method of high-frequency defect sensing complements the techniques for the detection of low-frequency TLS defects utilizing a spin-locking pulse sequence~\cite{Yan2013}. The key difference between those two approaches is that the reported method effectively probes a high-frequency transverse noise, while low-frequency spin-locking spectroscopy methods focus on a low-frequency longitudinal noise~\cite{Abdurakhimov2020}.    

In order to use the described approach for the identification of high-frequency TLS defects in fixed-frequency qubits, such as transmons, one can tune the qubit frequency by applying an in-plane magnetic field~\cite{VanDuzerTurner,BaronePaterno,Schneider2019,Krause2022} or by inducing an AC Stark shift using an off-resonant microwave drive~\cite{Carroll2022,Zhao2022}. If the qubit frequency tunability is completely unavailable, the reported technique cannot be used for the identification of a qubit-defect interaction type, but it can be applied to extract basic parameters of off-resonant TLS defects, including effective qubit-defect frequency detuning and coupling strength. 

Our results also imply that the always-on transverse coupling between a qubit and a high-frequency off-resonant TLS defect would coherently mix bare states of the coupled system. Even when the qubit and TLS defect are detuned, the time evolution of the qubit bare-state population would be characterized by fast small-amplitude oscillations, since the proper eigenstates of the hybrid system are not the bare states, but the qubit-defect entangled states. Those population oscillations due to the non-resonant (dispersive) interaction between the qubit and detuned TLS defects can result in additional gate and measurement errors~\cite{Galiautdinov2012,Matsuzaki2012,Khezri2015}. The importance of the reported method of TLS defect spectroscopy is that it allows one to map off-resonant TLS defects in a wide frequency band, and to extract defect parameters that can be used for numerical gate optimization.

\section{Theoretical model\label{theory}}

In this section, we present a theoretical model of the hybrid system composed of a high-frequency TLS defect and a qubit under a strong microwave drive. Here, a c-shunt flux qubit is modeled as a nonlinear Duffing oscillator similar to a transmon qubit. We derive qubit-defect interaction terms for different types of TLS defects and show that the shape of the interaction term determines the condition of the dynamical coupling between a qubit and an off-resonant TLS defect. 

\subsection{Qubit and TLS defect Hamiltonians}

We consider a c-shunt flux qubit consisting of a superconducting loop interrupted by three Josephson junctions shunted by a large capacitance $C_\text{S}$~\cite{You2007,Steffen2010,Yan2016,Abdurakhimov2019}. Two junctions are identical and characterized by the same critical current $I_\text{c}$ and capacitance $C_\text{J}$, while the area of the third junction is reduced by a factor of $\alpha$. The large shunt capacitance $C_\text{S} \gg C_\text{J}$ is connected in parallel with the smaller junction. In this case, the qubit Hamiltonian can be written in the one-dimensional form~\cite{Steffen2010,Yan2016,Abdurakhimov2019}:
\begin{equation}
\begin{split}
\hat{H}_\text{CSFQ} & = E_{C_\text{S}} \hat{n}^2 + 2E_\text{J}(1-\cos \hat{\varphi}) \\
& +\alpha E_\text{J} \left[1-\cos\left(2\pi\frac{\Phi_\text{e}}{\Phi_0}+2\hat{\varphi}\right)\right],
\end{split}
\label{csfq-H}
\end{equation}
where $E_{C_\text{S}} \approx e^2/2C_\text{S}$ is the effective charging energy, $E_\text{J}= \Phi_0 I_\text{c}/2\pi$ is the Josephson energy, $\hat{\varphi}$ and $\hat{n} = -i \partial / \partial \hat{\varphi}$ are phase and Cooper-pair number operators, respectively, $\Phi_\text{e}$ is an external magnetic flux, and $\Phi_0$ is a magnetic flux quantum. The effective charging energy $E_{C_\text{S}}$ is much smaller than the characteristic junction charging energy $E_\text{C} = e^2 /2 C_\text{J}$.

By taking into account charge and critical-current noise fluctuations relevant for the smaller Josephson junction (later it will be shown that those fluctuations are related to TLS defects located in this junction), the qubit Hamiltonian close to the optimal magnetic flux bias of $\Phi_\text{e}=0.5\Phi_0$ (``sweet spot'') can be written as
\begin{equation}
\begin{split}
& \hat{H}_\text{CSFQ} = E_{C_\text{S}}(\hat{n}+ \delta n)^2 + 2E_\text{J}(1-\cos \hat{\varphi}) \\
& + \alpha E_\text{J} \times \left( 1+ \frac{\delta I_\alpha}{\alpha I_\text{c}}\right) \times \left[1+\cos\left(2\hat{\varphi}+2\pi \delta f\right)\right],
\end{split}
\label{csfq-H-optimal-point}
\end{equation}
where $\delta n$ is the difference between electric charge fluctuations (in units of $2e$) on the two superconducting islands separated by the smaller Josephson junction~\cite{You2007}, $\delta I_\alpha$ is the fluctuation of the critical current through that junction, and $\delta f = (\Phi_e/\Phi_0) - 0.5$ is the relative flux detuning from the optimal magnetic flux bias.

The potential energy of the c-shunt flux qubit is determined by the second and third terms in the right-hand side of \cref{csfq-H-optimal-point}. In the case of $\alpha < 0.5$, the qubit has a single-well potential similar to a transmon qubit~\cite{Koch2007}. By expanding cosines for small angles and neglecting noise contributions, the qubit at the optimal magnetic flux bias can be treated as a quantum harmonic oscillator with a forth-order perturbation term, and the Hamiltonian can be written in the form of a Duffing oscillator~\cite{Yan2016,Abdurakhimov2019}:
\begin{equation}
    \hat{H}_\text{q} = \hbar\omega_b \left( \hat{b}^\dagger \hat{b} + \frac{1}{2}\right) + \frac{\hbar A}{12}\left( \hat{b}^\dagger + \hat{b} \right)^4,
\label{duffing}
\end{equation}
where $\omega_b$ is the characteristic frequency of the Duffing oscillator given by
\begin{equation}
\hbar\omega_b = \sqrt{4 E_{C_\text{S}} E_\text{J} (1-2\alpha)},
\end{equation}
and $A$ is the qubit anharmonicity given by
\begin{equation}
\hbar A = \frac{8\alpha-1}{4(1-2\alpha)}E_{C_\text{S}}.
\end{equation}
Here, $\hat{b}^\dagger$ and $\hat{b}$ are creation and annihilation bosonic operators that are related to the charge and phase operators by 
\begin{equation}
\hat{n} = \frac{i}{\sqrt{2}} \left[\frac{E_J(1-2\alpha)}{E_{C_\text{S}}}\right]^{1/4} \left( \hat{b}^\dagger - \hat{b} \right),
\label{eq-n}
\end{equation}
and 
\begin{equation}
\hat{\varphi} = \frac{1}{\sqrt{2}} \left[ \frac{E_{C_\text{S}}}{E_J(1-2\alpha)}\right]^{1/4} \left( \hat{b}^\dagger + \hat{b} \right).
\label{eq-varphi}
\end{equation}

Using the commutation relation $[\hat{b},\hat{b}^\dagger]=1$ and neglecting off-diagonal terms in~\cref{duffing} (these terms disappear under rotating wave approximation used below), the qubit Hamiltonian can be written as:
\begin{equation}
    \hat{H}_\text{Q}= \left(\hbar \omega_b + \frac{\hbar A}{2}\right) \hat{b}^\dagger \hat{b} +\frac{\hbar A}{2} \left(\hat{b}^\dagger \hat{b}\right)^2,
\label{eq-duffing-short}
\end{equation}
with the frequencies of the two lowest qubit transitions given by
\begin{equation}
\omega_\text{q} \equiv \omega_{01} = \omega_b + A,
\end{equation}
and
\begin{equation}
\omega_{12} = \omega_b + 2A.
\end{equation}
It can be seen from~\cref{eq-duffing-short} that, for a typical case when the qubit operational space is limited to the lowest two or three levels, the described perturbation approach should provide meaningful results when the condition $A \ll \omega_b$ is fulfilled.

The Hamiltonian of the TLS defect in the eigenstate basis is given by 
\begin{equation}
\hat{H}_\text{TLS}= \frac{1}{2} \hbar \omega_\text{TLS} \hat{\sigma}_z,
\end{equation}
where $\omega_\text{TLS}$ is the TLS defect frequency, and $\hat{\sigma}_z$ is the Pauli operator of the defect (details can be found in~\cref{appendix-defect-Hamiltonian}).

\subsection{Qubit-defect interaction}
The form of the qubit-defect interaction term $H_\text{int}$ depends on a particular microscopic mechanism of the coupling between a qubit and a TLS defect~\cite{Muller2019}. In the most commonly used model, it is assumed that a charge TLS defect induces the charge fluctuations $\delta n_\text{TLS}$ across a relevant Josephson junction~\cite{Martinis2005,Muller2019}, and the corresponding interaction Hamiltonian can be obtained by substituting $\delta n = \frac{1}{2} \delta n_\text{TLS} \hat{\sigma}_x$ into \cref{csfq-H-optimal-point}:
\begin{equation}
\hat{H}_\text{int}^{\text{(C)}} = i \hbar g_\text{C} \hat{\sigma}_x\left( \hat{b}^\dagger - \hat{b} \right),
\label{charge-defect}
\end{equation}
where $\hat{\sigma}_x$ is the Pauli operator of the TLS defect, and $g_\text{C}$ is the coupling strength between the qubit and defect given by:
\begin{equation}
g_\text{C} = \frac{1}{2} \delta n_\text{TLS} \sqrt{\frac{\omega_b E_{C_\text{S}}}{\hbar}}.
\end{equation}
According to \cref{charge-defect}, the coupling between a qubit and a standard charge TLS defect is linear in terms of the qubit operators $\hat{b}^\dagger$ and $\hat{b}$. 

Another possible interaction mechanism between a qubit and a TLS defect is related to critical-current fluctuations across a Josephson junction~\cite{Simmonds2004,Muller2019}. Since the thickness of the junction tunnel barrier is non-uniform, the tunneling of Cooper pairs occurs through a discrete set of conductance channels. Fluctuations in the charge configuration of a TLS defect can block one of the conductance channels, resulting in the fluctuations of the critical current $\delta I_\text{TLS}$ through that junction~\cite{Muller2019}. Other possible microscopic models of critical-current noise include Andreev fluctuators~\cite{Faoro2005,deSousa2009}, and Kondo-like traps~\cite{Faoro2007}. The interaction Hamiltonian due to critical-current fluctuations can be estimated by substituting $\delta I_\alpha = \frac{1}{2}\delta I_\text{TLS} \hat{\sigma}_x$ into~\cref{csfq-H-optimal-point}. By expanding the cosine term $\cos(2\hat{\varphi}+2\pi \delta f)$ for the small phase $\hat{\varphi}$ and flux detuning $\delta f$, the interaction term can be written in terms of qubit operators $\hat{b}^\dagger$ and $\hat{b}$ as:
\begin{equation}
\hat{H}_\text{int} = \hbar g_I^{(1)} \hat{\sigma}_x \left( \hat{b}^\dagger + \hat{b} \right) + \hbar g_I^{(2)} \hat{\sigma}_x \left( \hat{b}^\dagger + \hat{b} \right)^2,
\end{equation}
where coupling strengths $g_I^{(1)}$ and $g_I^{(2)}$ are given by:
\begin{equation}
g_I^{(1)} = -  \pi \delta f \frac{\alpha}{\sqrt{1-2\alpha}}  \left( \frac{\delta I_\text{TLS}}{\alpha I_\text{c}} \right) \sqrt{\frac{\omega_b E_J}{\hbar}},
\end{equation}
and
\begin{equation}
g_I^{(2)} = -  \frac{\alpha}{4(1-2\alpha)}  \left( \frac{\delta I_\text{TLS}}{\alpha I_\text{c}} \right) \omega_b .
\label{g_I}
\end{equation}

In this work, it is assumed that the qubit is operated close to the optimal flux bias point ($\delta f \approx $ 0), and, hence, the interaction between the qubit and TLS defect through critical-current fluctuations is described by the nonlinear term given by:
\begin{equation}
\hat{H}_\text{int}^{\text{(I)}} = \hbar g_I^{(2)} \hat{\sigma}_x \left( \hat{b}^\dagger + \hat{b} \right)^2.
\label{current-defect}
\end{equation}

\subsection{Qubit-defect system under a strong drive: linear coupling}
\begin{figure}[t!]
    \centering
    \includegraphics{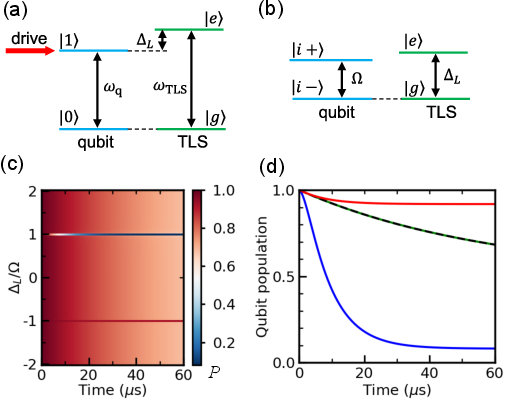}
    \caption{(a,b) Energy level diagrams of a qubit and an off-resonant charge-fluctuation TLS defect (a) in the laboratory frame, and (b) in the rotating frame for $\Delta_\text{L}>0$. (c,d) Results of numerical simulations of the system evolution for a given Rabi frequency: (c) The population of the excited qubit state $\ket{i+}$ in the rotating frame as a function of the evolution time and qubit-defect detuning.  The qubit is dynamically coupled to the defect when the condition given by~\cref{condition-linear} is met. (d) The qubit population at the qubit-defect detunings $\Delta_L = \Omega$ (blue line), $\Delta_L = -\Omega$ (red line), and $\Delta_L \neq \pm \Omega$ (green line). The dashed black line corresponds to the exponential decay $f(t)=[1+\exp{(-\Gamma_{1\text{q}} t/2})]/2$.}
    \label{fig1}
\end{figure}

First, we consider the driven evolution of the qubit coupled to a standard charge TLS defect. We assume that there is some detuning between the qubit and defect frequencies as shown in~\cref{fig1}(a). The system Hamiltonian is given by
\begin{equation}
\hat{H} = \hat{H}_\text{Q} + \hat{H}_\text{TLS} + \hat{H}_\text{int}^{\text{(C)}} + \hat{H}_\text{d}.
\end{equation}
The driving term $\hat{H}_\text{d}$ is given by
\begin{equation}
\hat{H}_\text{d} = i \hbar \Omega \cos(\omega_\text{q} t) \left( \hat{b}^\dagger - \hat{b} \right),
\label{driving-term}
\end{equation}
where $\Omega$ is the Rabi angular frequency. Here, it is assumed that the qubit is capacitively coupled to a readout cavity mode via the charge degree of freedom, and the microwave drive is applied at the qubit resonance frequency.

Using rotating wave approximation (RWA) (details can be found in~\cref{RWA}), the system Hamiltonian in the rotating frame is given by 
\begin{equation}
\begin{split}
\hat{H}^\text{R} = & \frac{\hbar A}{2} \left[ (\hat{b}^\dagger \hat{b})^2 - \hat{b}^\dagger \hat{b} \right] + i\frac{\hbar \Omega}{2} (\hat{b}^\dagger - \hat{b}) + \frac{\hbar \Delta_\text{L}}{2}\hat{\sigma}_z \\
                  & + i\hbar g_\text{C} (\hat{b}^\dagger\hat{\sigma}_- - \hat{b} \hat{\sigma}_+). 
\end{split}
\end{equation}
where $\hat{\sigma}_\pm = \frac{1}{2}(\hat{\sigma}_x \pm i \hat{\sigma}_y)$, and the separation between defect energy levels $\Delta_\text{L}$ is given by
\begin{equation}
\Delta_\text{L} = \omega_\text{TLS} - \omega_\text{q}.
\end{equation} 

By truncating the qubit state space to the lowest two states, the qubit eigenstate basis for the non-interacting part of the Hamiltonian in the rotating frame is given by
\begin{equation}
\ket{i-} = \frac{\ket{0}-i\ket{1}}{\sqrt{2}},
\end{equation}
and
\begin{equation}
\ket{i+} = \frac{\ket{0}+i\ket{1}}{\sqrt{2}}.
\end{equation}

In the $\{\ket{i-},\ket{i+}\}$ eigenstate basis, the system Hamiltonian is given by
\begin{equation}
\hat{H}^\text{R} = -\frac{\hbar \Omega}{2} \hat{s}_z + \frac{\hbar \Delta_\text{L}}{2}\hat{\sigma}_z - \frac{\hbar g_\text{C}}{2} (\hat{s}_z\hat{\sigma}_x + \hat{s}_y \hat{\sigma}_y),
\label{charge-RWA} 
\end{equation}
where $\hat{s}_{x,y,z}$ are the qubit Pauli operators. By performing a second RWA transformation, it can be shown that the interaction term proportional to $\hat{s}_z\hat{\sigma}_x$ can be neglected, and, hence, the effective coupling strength is equal to $g_\text{C}/2$. The diagram of energy levels of the qubit and defect in the rotating frame is shown in~\cref{fig1}(b).

We performed numerical simulations of the system evolution by solving a Lindblad master equation for the Hamiltonian~\cref{charge-RWA} in the QuTiP package~\cite{Johansson2012,Johansson2013} [\cref{fig1}(c,d)]. The collapse operators were chosen in the form of $\hat{C}_1 = \sqrt{\Gamma_{1\text{q}}} \hat{b} = \sqrt{\Gamma_{1\text{q}}} ( - \hat{s}_y -i \hat{s}_z)/2$ and $\hat{C}_2=\sqrt{\Gamma_{1\text{TLS}}}\hat{\sigma}_-$. Here, the qubit pure dephasing is not taken into account, since, for sufficiently high Rabi frequencies, the driven qubit state should be effectively decoupled from low-frequency dephasing noises such as a $1/f$ phase noise~\cite{Abdurakhimov2020}. The qubit was initialized in its excited state $\ket{i+}$ in the rotating frame, and the defect was initialized in its ground state $\ket{g}$. The simulation parameters were $\Omega/2\pi=$\,25\,MHz, $g_\text{C}/2\pi=$\,50\,kHz, $\Gamma_{1\text{q}}=$\,0.03\,$\mu$s$^{-1}$, and $\Gamma_{1\text{TLS}}=$\,1\,$\mu$s$^{-1}$.

The condition for the dynamical coupling between a qubit and a defect with a linear coupling to the qubit is given by:
\begin{equation}
\Omega = | \Delta_L | = |\omega_\text{TLS}-\omega_\text{q}|.
\label{condition-linear}
\end{equation}
If the driven qubit state is not coupled to the defect, i.e., $\Omega \neq | \Delta_L |$, the qubit relaxation rate in the rotating frame is given by $\Gamma_{1\rho}=\Gamma_{1\text{q}}/2$~[\cref{fig1}(d)]. When the condition~\cref{condition-linear} is fulfilled, the relaxation of the driven qubit state is affected by the interaction with the defect, and, in the case $\Gamma_{1\text{q}} < g_\text{C} < \Gamma_{1\text{TLS}} $, the effective relaxation rate of the qubit can be roughly estimated by a Purcell-like formula, $\Gamma_\text{P}\approx g_\text{C}^2/\Gamma_{1\text{TLS}}$~\cite{Abdurakhimov2020}. The sign of the offset of the stationary qubit population level from the value of 0.5 depends on the sign of the qubit-defect detuning $\Delta_L$~[\cref{fig1}(d)], and, hence, one can say that spectral signatures for defects with negative and positive detunings have different ``polarities''.
Depending on the ``polarity'' (i.e., the sign of the detuning $\Delta_L$), a TLS defect can be considered as  a ``cold'' or a ``hot'' subsystem~\cite{Abdurakhimov2020}. When the condition $\Delta_L > 0$ is fulfilled, the ground state of the TLS defect in the laboratory frame corresponds to the ground state in the rotating frame~[\cref{fig1}(b)], and the population is transferred from the qubit to the ``cold'' TLS defect. In contrast, when the condition $\Delta_L < 0$ is met, the ground state of the TLS defect in the laboratory frame corresponds to the excited state in the rotating frame, and the population is transferred from the ``hot'' TLS defect to the qubit (details can be found in~\cref{hot-defect}).

\begin{figure}[t!]
    \centering
    \includegraphics{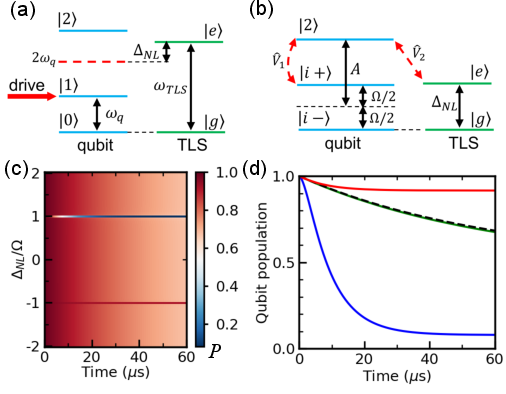}
    \caption{(a,b) Schematics of energy levels of a qubit and a critical-current defect (a) in the laboratory frame, and (b) in the rotating frame for $\Delta_\text{NL}>0$. Here, $\hat{V}_1$ and $\hat{V}_2$ correspond to the interaction terms described in~\cref{V1} and~\cref{V2}, respectively. (c,d) Results of numerical simulations of the system evolution for a given Rabi frequency in the case of a critical-current-fluctuation TLS defect: (c) The population of the qubit state $\ket{i+}$ as a function of the evolution time and qubit-defect detuning.  The qubit is dynamically coupled to the defect when the condition given by~\cref{condition-nonlinear} is met. (d) The qubit population at the qubit-defect detunings $\Delta_{NL} = \Omega$ (blue line), $\Delta_{NL} = -\Omega$ (red line), and $\Delta_{NL} \neq \pm \Omega$ (green line). The dashed black line corresponds to the exponential decay $f(t)=[1+\exp{(-\Gamma_{1\text{q}} t/2})]/2$.}
    \label{fig2}
\end{figure}

\subsection{Qubit-defect system under a strong drive: nonlinear coupling}

In this section, we consider the driven evolution of the qubit coupled to a critical-current-fluctuation TLS defect. The system Hamiltonian is given by
\begin{equation}
\hat{H} = \hat{H}_\text{Q} + \hat{H}_\text{TLS} + \hat{H}_\text{int}^{\text{(I)}} + \hat{H}_\text{d}.
\label{Hamiltonian-current-full}
\end{equation}

By using RWA (details can be found in~\cref{RWA}), the Hamiltonian can be written in the rotating frame as
\begin{equation}
\begin{split}
\hat{H}^\text{R} = & \frac{\hbar A}{2} \left[ (\hat{b}^\dagger \hat{b})^2 - \hat{b}^\dagger \hat{b} \right] + i\frac{\hbar \Omega}{2} (\hat{b}^\dagger - \hat{b}) + \frac{\hbar \Delta_{NL}}{2}\hat{\sigma}_z \\
                  & + \hbar g_I^{(2)} (\hat{b}^{\dagger 2}\hat{\sigma}_- + \hat{b}^2 \hat{\sigma}_+), 
\end{split}
\label{Hamiltonian-current}
\end{equation}
where the detuning is defined as
\begin{equation}
\Delta_{NL} = \omega_\text{TLS} - 2 \omega_\text{q}.
\end{equation}

We performed numerical simulations of the system evolution by solving the master equation for the Hamiltonian~\cref{Hamiltonian-current}.
The qubit states were truncated to the lowest three levels. The collapse operators were $\hat{C}_1 = \sqrt{\Gamma_{1\text{q}}} \hat{b}$ and $\hat{C}_2=\sqrt{\Gamma_{1\text{TLS}}}\hat{\sigma}_-$. The qubit was initialized in its excited state $\ket{i+}$ in the rotating frame, and the defect was initialized in its ground state $\ket{g}$. The simulation parameters were $\Omega/2\pi=$\,25\,MHz, $A/2\pi=$\,1\,GHz, $g_I^{(2)}/2\pi=$\,2\,MHz, $\Gamma_{1\text{q}}=$\,0.03\,$\mu$s$^{-1}$, and $\Gamma_{1\text{TLS}}=$\,1\,$\mu$s$^{-1}$. The population of the qubit state $\ket{i+}$ is shown in~\cref{fig2}(c,d).

The condition for the dynamical coupling between a qubit and a defect with a nonlinear coupling to the qubit is determined by:
\begin{equation}
\Omega = | \Delta_{NL} | = |\omega_\text{TLS}-2\omega_\text{q}|,
\label{condition-nonlinear}
\end{equation}
which is different from~\cref{condition-linear} for a standard charge-fluctuation defect.

The numerical results imply that the qubit can be coupled to a critical-current defect located close to the double qubit frequency $2\omega_\text{q}$ as shown in~\cref{fig2}(a).  
We further clarify the mechanism of such coupling by rewriting the Hamiltonian given by~\cref{Hamiltonian-current} in the qubit basis $\{\ket{i-}, \ket{i+},\ket{2}\}$ and defect basis $\{\ket{g},\ket{e}\}$ in the rotating frame: 
\begin{equation}
\hat{H}^\text{R} = \hat{H}_0 + \hat{V}_1 + \hat{V}_2,
\end{equation}
where the non-interacting term $\hat{H}_0$ is given by
\begin{equation}
\begin{split}
\hat{H}_0  & = \frac{\hbar \Omega}{2} \left( \ket{i+} \bra{i+} -\ket{i-} \bra{i-} \right) + \hbar A \ket{2}\bra{2} \\
      & + \frac{\hbar \Delta_{NL}}{2} \hat{\sigma}_z, 
\end{split}
\end{equation}
and the interaction terms $\hat{V}_1$ and $\hat{V}_2$ are given by
\begin{equation}
\hat{V}_1 = \frac{\hbar \Omega}{2} \left(\ket{i-} \bra{2} - \ket{i+} \bra{2}  + \text{h.c.} \right),
\label{V1}
\end{equation}
and
\begin{equation}
\hat{V}_2 = \hbar g_I^{(2)} \left( \ket{2,g} \bra{i+,e} + \ket{2,g} \bra{i-,e} + \text{h.c.} \right).
\label{V2}
\end{equation}
Using the second-order perturbation theory~\cite{Kockum2017}, the effective coupling strength between the states $\ket{i+,g}$ and $\ket{i-,e}$ is given by
\begin{equation}
g_\text{eff} = \frac{\braket{i-,e|\hat{V}_2|2,g}\braket{2,g|\hat{V}_1|i+,g}}{\hbar^2 (\frac{\Omega}{2}-A)} \approx \frac{g_I^{(2)} \Omega}{2A},
\label{effective-coupling-strength}
\end{equation}
and, thus, the interaction between the qubit and defect is mediated by virtual transitions via the qubit second excited state as shown in~\cref{fig2}(b). 

An equation similar to the condition~\cref{condition-nonlinear} can be obtained by considering counter-rotating interaction terms omitted in~\cref{Hamiltonian-current}, but the corresponding coupling strength would be smaller than the one given by~\cref{effective-coupling-strength} (details can be found in \cref{counter-rotating-coupling}).

\begin{figure*}[t!]
    \centering
    \includegraphics[width=\textwidth]{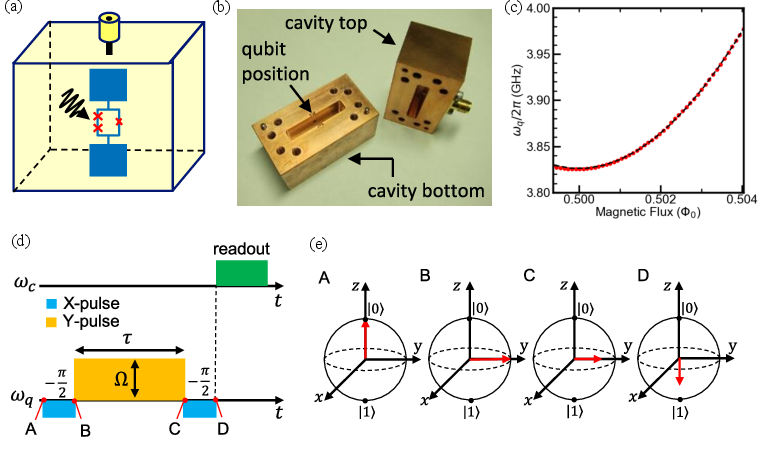}
    \caption{(a) The schematic of a c-shunt flux qubit embedded inside a 3D microwave cavity. (b) The photograph of the 3D microwave cavity. (c) The qubit spectrum as a function of an applied magnetic flux in the units of the magnetic flux quantum $\Phi_0$. Red dots represent the experimental data. The error bars are within the dot size (the frequency fitting error was less than 0.1 MHz). The dashed black line corresponds to results of numerical simulations using the scqubits package. (d) The schematics of microwave pulse sequences used for the cavity readout (top) and qubit drive (bottom). (e) The Bloch-sphere representation of qubit state vectors at different stages of the spin-locking pulse sequence A, B, C, and D shown in~\cref{fig3}(d).}
    \label{fig3}
\end{figure*}

\section{Experimental demonstration}

The reported method was experimentally demonstrated by performing TLS defect spectroscopy in a c-shunt flux qubit coupled to a 3D microwave cavity~[\cref{fig3}(a),(b)]. The 3D microwave cavity provided a clean electromagnetic environment without spurious microwave modes, which was important for reliable identification of TLS defects. The qubit was fabricated on a high-resistivity silicon substrate by double-angle shadow evaporation of aluminum. Details of the qubit design can be found in the previous work~\cite{Abdurakhimov2019}. The qubit was measured via dispersive readout using the experimental setup described in~\cref{appendix-measurement-setup}. The qubit spectrum as a function of the applied magnetic flux $\Phi_e$ is shown in~\cref{fig3}(c). The area of the qubit loop was about 16\,$\mu$m$^2$, and, hence, the magnetic flux bias of $0.5\Phi_0$ corresponded to the applied magnetic field of approximately 65\,$\mu$T. By fitting the spectrum using the scQubits Python package~\cite{Groszkowski2021}, the parameters of the c-shunt flux qubit were estimated to be $\alpha \approx 0.457$, $E_\text{C}/h \approx $\,3.2\,GHz, $E_{C_\text{S}}/h \approx $\,0.24\,GHz, and $E_\text{J}/h \approx $\,160\,GHz. In separate measurements at the optimal flux bias point of $\Phi_e = 0.5\Phi_0$, the following system parameters were determined: cavity resonance frequency $\omega_c/2\pi \approx $~8.192\,GHz, qubit transition frequency $\omega_\text{q}/2\pi \approx $~3.825\,GHz, qubit anharmonicity $A/2\pi \approx$~1\,GHz, qubit energy-relaxation time $T_1\approx$~53\,$\mu$s, and qubit Hahn-echo dephasing time $T_\text{2E}\approx$~34\,$\mu$s. 

We detected TLS defects by measuring the qubit excited-state population immediately after the application of a strong microwave drive at the qubit resonance frequency $\omega_q$. The qubit was driven by a so-called spin-locking pulse sequence, where a strong Y-pulse of the duration $\tau$ and amplitude $\Omega$ was preceded and followed by low-amplitude X-pulses corresponding to $-\pi/2$ rotations of the qubit state vector~[\cref{fig3}(d)]. As shown in~\cref{fig3}(e), the first pulse rotated the qubit state vector  around the X-axis  from the initial $\ket{0}$ state to the $\ket{i+}$ state aligned along the Y-axis. The second pulse was a strong microwave Y-pulse that rotated the $\ket{i+}$ state around the Y-axis and generated the required driving term~[\cref{driving-term}]. The third pulse rotated the qubit state to its final state oriented along the Z-axis, and the final qubit population was measured by dispersive readout. Thus, the spin-locking pulse sequence allowed us to effectively convert between the qubit states in the basis of $\{\ket{0},\ket{1}\}$, which was preferable for dispersive readout, and the qubit states in the basis of $\{\ket{i-},\ket{i+}\}$, which was an eigenstate basis of the strongly driven qubit as described in~\cref{theory}.

\begin{figure*}[t!]
    \centering
    \includegraphics[width=\textwidth]{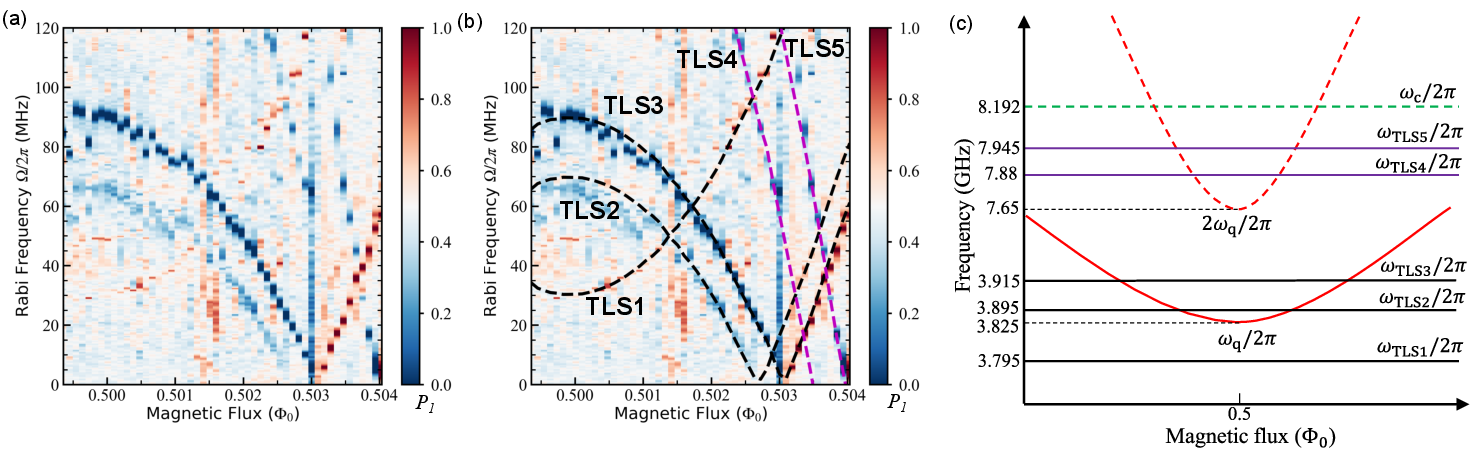}
    \caption{Experimental demonstration of TLS defect detection using a strong qubit drive. (a) The final exited-state qubit population as a function of the applied magnetic flux and the qubit drive amplitude in Rabi-frequency units. (b) Results of the fitting of the experimental data. Spectral lines TLS1, TLS2, and TLS3 can be fit by~\cref{condition-linear} that corresponds to charge-fluctuation TLS defects.  Spectral lines TLS4 and TLS5 can be fit by~\cref{condition-nonlinear} that corresponds to critical-current-fluctuation TLS defects. (c) The energy level diagram of the system (not to scale). Solid and dashed red lines correspond to the qubit transition frequency $\omega_\text{q}$ and the double value of  $\omega_\text{q}$, respectively. Black and magenta horizontal lines represent charge-fluctuation and critical-current-fluctuation TLS defects, respectively. The dashed green line corresponds to the cavity resonance frequency.}
    \label{fig4}
\end{figure*}

Figure~\ref{fig4}(a) shows results of the measurement of the qubit excited-state population as a function of the applied magnetic flux and amplitude of the spin-locking Y-pulse. Additional time-domain data are shown in~\cref{preliminary}, which correspond to simulations shown in~\cref{fig1}(c) and~\cref{fig2}(c). In experiments, we recalibrated all flux-dependent measurement parameters --- including the cavity frequency, qubit frequency, Rabi frequency, and the corresponding duration of an X-pulse --- at each applied magnetic flux value. The signal-to-noise ratio was improved by using the phase-cycling method~(\cref{phase-cycling}). The amplitude of the spin-locking Y-pulse was calibrated in the units of the Rabi frequency $\Omega$~(\cref{calibration}).  The appropriate duration of the spin-locking Y-pulse $\tau =$\,60\,$\mu$s was determined in time-domain measurements~(\cref{preliminary}). The amplitudes of X-pulses were fixed (in voltage units), while their duration was adjusted at each magnetic flux bias to ensure the required rotation angle. Typically, the X-pulse duration was about 40 ns. The repetition period of the sequence was sufficiently long (typically, about 1\,ms) to keep the temperature of the mixing chamber stage of the dilution refrigerator below 30\,mK.

\begin{figure}[t!]
    \centering
    \includegraphics[width=\columnwidth]{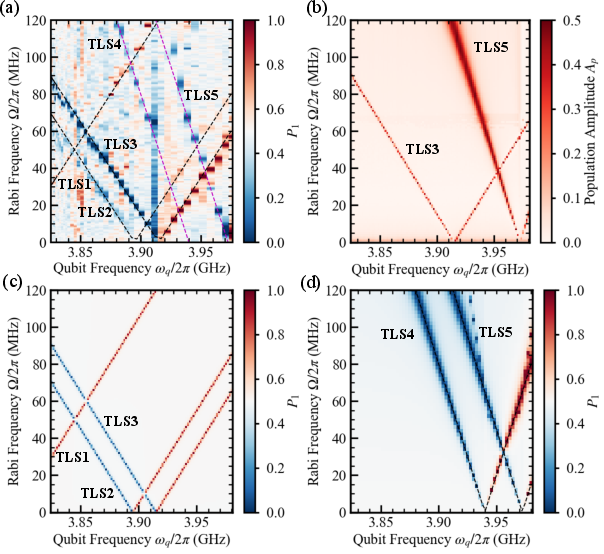}
    \caption{(a) The experimental data shown in~\cref{fig4}(a,b) plotted as a function of the qubit frequency~$\omega_q$. Straight dashed lines represent analytic equations described in the text. (b)~Results of numerical simulations of the system evolution using a time-dependent system Hamiltonian without RWA. The system consisted of the qubit, charge defect TLS3, and critical-current defect TLS5 (other defects were not included in the numerical model in order to minimize the calculation time). The amplitude of qubit population oscillations was calculated using the procedure described in the~\cref{time-depend}. (c,d)~Results of numerical calculations of the hybrid system evolution using RWA. The system consisted of the qubit coupled to (c)~charge-fluctuation defects TLS1--TLS3, and (d)~critical-current-fluctuation defects TLS4 and TLS5, described by~\cref{charge-RWA} and~\cref{Hamiltonian-current}, respectively.}
    \label{fig5}
\end{figure}

Pronounced spectral lines TLS1--TLS5 were observed in the experimental data~[\cref{fig4}(b) and \cref{fig5}(a)]. Spectral lines TLS1, TLS2, and TLS3 were fit by~\cref{condition-linear} with the TLS defect frequencies $\omega_\text{TLS1}/2\pi \approx $\,3.795\,GHz, $\omega_\text{TLS2}/2\pi \approx $\,3.895\,GHz, and $\omega_\text{TLS3}/2\pi \approx $\,3.915\,GHz. Thus, the spectral lines TLS1--TLS3 were due to the interactions between the qubit and conventional charge-fluctuation TLS defects. Since the qubit frequency was greater than the frequency $\omega_\text{TLS1}$ in the whole range of the applied magnetic flux bias, the qubit-defect detuning was negative, $\Delta_L < 0$, and, hence, the spectral line TLS1 was described by the equation $\Omega = \omega_q - \omega_\text{TLS1}$. As for the TLS defects corresponding to the spectral lines TLS2 and TLS3, the qubit frequency was less than the frequencies $\omega_\text{TLS2}$ and $\omega_\text{TLS3}$ near the optimal flux bias point, but the qubit transition frequency crossed the defect levels at the flux bias values close to $\Phi_e  \approx 0.503 \Phi_0$. Therefore, spectral lines TLS2 and TLS3 could be fit by equations  $\Omega = \omega_\text{TLS2}-\omega_q$ and $\Omega = \omega_\text{TLS3}-\omega_q$ in the vicinity of the optimal point, and by equations  $\Omega = \omega_q - \omega_\text{TLS2}$ and $\Omega = \omega_q - \omega_\text{TLS3}$ at magnetic flux values $\Phi_e   \gtrsim 0.503 \Phi_0$, respectively. The interaction between the qubit and TLS defects with positive (negative) values of the defect-qubit detuning, $\Delta_L> 0$ ($\Delta_L < 0$), resulted in the formation of local minima (maxima) in the qubit population signal. Thus, spectral lines TLS2 and TLS3 had two different ``polarities'' depending on the applied magnetic flux bias, and they changed their ``polarity'' near the flux bias $\Phi_e = 0.503 \Phi_0$, where the corresponding TLS defects had $\Delta_L=0$. This behavior was consistent with the predictions of the theoretical model described in~\cref{theory}. No avoided crossing was observed in the qubit spectrum near $\Phi_e  \approx 0.503 \Phi_0$, but the qubit Rabi oscillations were suppressed due to the resonant qubit-defect interaction. Therefore, at that particular magnetic flux value, the automatic procedure of the measurement parameter recalibration did not provide correct values, resulting in the appearance of the vertical line at $\Phi_e  \approx 0.503 \Phi_0$ in~\cref{fig4}(a,b). By fitting the time-domain spin-locking data~(\cref{preliminary}), we estimated the values of the charge-fluctuation coupling strength $g_\text{C}/2\pi\approx$\,50\,kHz and defect relaxation rate $\Gamma_{1\text{TLS}}\approx $\,1\,$\mu$s$^{-1}$ which were consistent with the values reported in the previous works~\cite{Burnett2019,Lisenfeld2019,Abdurakhimov2020}.

In contrast to the spectral features of TLS1--TLS3, positions of spectral lines TLS4 and TLS5 could not be described by~\cref{condition-linear}. Instead, the spectral signatures of TLS4 and TLS5 were fit by~\cref{condition-nonlinear} with TLS defect frequencies $\omega_\text{TLS4}/2\pi \approx $\,7.88\,GHz and $\omega_\text{TLS5}/2\pi \approx $\,7.945\,GHz, respectively. Thus, spectral lines TLS4 and TLS5 were formed due to the interaction between the qubit and critical-current-fluctuation defects. By fitting the experimental data obtained in time-domain spin-locking measurements at a fixed magnetic flux bias~(\cref{preliminary}), the typical coupling strength between the qubit and TLS5 defect  was estimated to be in the range of $g_I^{(2)}/2\pi \approx $\,8--22\,MHz, corresponding to the effective coupling strength of up to~1\,MHz according to~\cref{effective-coupling-strength}. Using~\cref{g_I}, the relative critical-current fluctuation $r = \delta I_\text{TLS}/ \alpha I_{c}$ was calculated to be in the range of $ r \approx$\,0.002--0.006. Based on the measurements of the $1/f$ critical-current noise in Josephson junctions at low frequencies, the value of the relative critical-current fluctuation due to a single TLS defect was estimated to be $r_0 \approx 10^{-4}$ for an aluminum-oxide junction with the area of 0.08\,$\mu$m$^2$~\cite{Simmonds2004}. In our case, the area of the small junction was about 0.01\,$\mu$m$^2$, and, by scaling the value of $r_0$ by the ratio of the junction areas, we obtained $r_0 \approx 8 \times 10^{-4}$ for our qubit which was close to the values $r$ estimated from the experimental data. Here, we assumed that the qubit was coupled to a single TLS defect (if the qubit was coupled to an ensemble of TLS defects, the effective coupling strength would scale with the square root of the total defect number). It should be also noted that the value of $r_0$ can depend on the junction fabrication technology.

The experimental results shown in~\cref{fig4}(a,b) can be also plotted as a function of the qubit frequency $\omega_q$ and the drive amplitude $\Omega$~[\cref{fig5}(a)]. Here, the conversion between the applied flux bias and qubit frequency was performed using the qubit spectrum data $\omega_q(\Phi_e)$ presented in~\cref{fig3}(c).
Experimental results can be reproduced well by numerical simulations of the driven qubit evolution as a function of the qubit frequency~[\cref{fig5}(b-d)].
Here,~\cref{fig5}(b) shows results of calculations using a time-dependent system Hamiltonian without RWA (details can be found in~\cref{time-depend}). Using that approach, it is possible to model all types of TLS defects simultaneously, but with the drawback of a long computation time. Simulations can be performed more efficiently using RWA, but, in this case, it is necessary to simulate each type of a TLS defect separately, since different types of TLS defects require different RWA transformations (details can be found in~\cref{RWA}). For example,~\cref{fig5}(c,d) show results of separate numerical simulations of charge-fluctuation and critical-current fluctuation TLS defects using ~\cref{charge-RWA} and~\cref{Hamiltonian-current}, respectively. In the numerical calculations, the coupling strengths for charge defects and critical-current defects were $g_\text{C}/2\pi\approx$\,100\,kHz and $g_I^{(2)}/2\pi = $\,20\,MHz, respectively, and the relaxation rates were the same for all TLS defects, $\Gamma_{1\text{TLS}}\approx $\,1\,$\mu$s$^{-1}$. As shown in~\cref{fig5}(b,d), spectral lines corresponding to the critical-current-fluctuation defects became less pronounced at small drive amplitudes $\Omega$ which was in qualitative agreement with the analytic prediction given by~\cref{effective-coupling-strength}. We were also able to reproduce different signal ``polarities'' in~\cref{fig5}(c,d). Clearly, our theoretical model accounts for experimental findings.

It was previously shown that detrimental effects of TLS defects can be mitigated by saturating TLS defects using a direct microwave excitation~\cite{Andersson2021,Niepce2021}, or by utilizing a qubit to heat or cool its TLS environment via a dynamical polarization effect~\cite{Spiecker2022}. In our experiments, a strong microwave drive was applied at the qubit transition frequency, and, hence, TLS defects were not excited directly. As for the dynamical polarization method, the qubit-defect interaction rates were much smaller than the energy-relaxation rates of TLS defects, and, therefore, we could not saturate TLS defects even at high drive amplitudes.

We repeated TLS spectroscopy measurements with an applied in-plane magnetic field of about 0.2\,mT~(\cref{magnetic-field}). No dependence of spectral line positions on the applied magnetic field was found, and, hence, the observed TLS defects were charge defects. It was also found that defect frequencies slightly drifted on a time scale of days during the same cooldown, and spectral distributions of TLS defects were different for different cooldowns of the same device.

\section{Discussion}
Building a realistic noise model of quantum processors is crucial for the progress of NISQ computing~\cite{Preskill2018,Cao2021,Bharti2022} and performance of quantum error correction (QEC) protocols with biased-noise superconducting qubits~\cite{Aliferis2009,Stephens2013,Tuckett2018,Tuckett2019}. For example, in the models of highly biased noise, it is usually assumed that $Z$ errors (dephasing) occur much more frequently than $X$ and $Y$ errors (energy relaxation). The presence of an off-resonant parasitic TLS defect can increase the rates of $X$ and $Y$ errors, since transverse qubit-defect interaction terms given by~\cref{charge-defect} and~\cref{current-defect} provide additional channels for energy relaxation. Therefore, a thorough characterization of off-resonant TLS defects is important for the determination of the dominant type of noise errors for a given qubit drive amplitude. The presented method of TLS defect spectroscopy provides information about the spectral distribution of off-resonant TLS defects, and it allows one to determine the exact form of qubit-defect interaction which is not accessible using other experimental techniques.

The reported technique allowed us to consistently distinguish between charge-fluctuation and critical-current-fluctuation TLS defects. Although high-frequency critical-current-fluctuation defects were discussed previously in the relation to experiments with phase qubits~\cite{Simmonds2004}, it was later shown that those results were better described by charge fluctuations~\cite{Martinis2005}. Regarding the interaction between a qubit and a critical-current-fluctuation defect, the key difference between a phase qubit and the c-shunt flux qubit used in this work is that a phase qubit is typically biased near the critical current of its Josephson junction, where  the superconducting phase difference across the Josephson junction $\varphi$ is close to $\pi/2$~\cite{Simmonds2004}. In that case, the coupling term between a phase qubit and a critical-current TLS defect is linear and proportional to $\cos{\varphi} \sim \tilde{\varphi}$, where $\tilde{\varphi} = \varphi - \pi/2$. In contrast, in the case of the c-shunt flux qubit biased near the optimal point $\Phi_e = 0.5 \Phi_0$, the effective phase $\varphi$ was small, $\varphi \approx 0$, and the qubit was coupled to critical-current defects via a nonlinear term proportional to $\cos{\varphi} \sim \varphi^2$. This nonlinearity allowed us to reliably distinguish critical-current TLS defects from standard charge TLS defects with a linear coupling. The reported results demonstrate that critical-current noise is particularly relevant for capacitively-shunted qubits, for which the characteristic Josephson energy $E_\text{J}$ is large, while the charge noise is suppressed by the large shunt capacitance, $E_\text{J} \gg E_{C_\text{S}}$. It should be noted that a nonlinear coupling to critical-current defects should be present in other types of qubits, including fixed-frequency transmons~\cite{Koch2007} and flux-tunable SNAIL transmons~\cite{Frattini2017,Grimm2020}. The method described in this work can be used for testing new materials and fabrication techniques that aim at minimizing the number of TLS defects in superconducting qubits, such as Josephson-junction fabrication based on epitaxial trilayer structures~\cite{Oliver2013,Kim2021}.

Our work implies that off-resonant high-frequency TLS defects can significantly affect the dynamics of a superconducting qubit. In the simplest case of a single charge-fluctuation TLS defect, the single-excitation subspace $\{\ket{1g},\ket{0e}\}$ is coherently mixed due to the always-on transverse qubit-defect coupling, and the eigenstates of the coupled system can be approximated by entangled states~\cite{Khezri2015}:
\begin{equation}
\begin{split}
\overline{\ket{1g}} & = \sqrt{1-\left(\frac{g}{\Delta}\right)^2}\ket{1g} - \frac{g}{\Delta}\ket{0e}, \\
\overline{\ket{0e}} & =\sqrt{1-\left(\frac{g}{\Delta}\right)^2}\ket{0e} + \frac{g}{\Delta}\ket{1g},
\end{split}
\end{equation}
where $g$ and $\Delta = \omega_\text{TLS}-\omega_\text{q}$ are qubit-defect coupling strength and frequency detuning, respectively, and it is assumed that the system is in the  dispersive regime, $g \ll \Delta$. If the measurement process occurs in the qubit bare state basis, the state $\overline{\ket{1g}}$ can be ``erroneously'' measured as $\ket{0e}$ with a probability of $(g/\Delta)^2$, which results in additional measurement error on the order of $(g/\Delta)^2$ depending on the details of the measurement process~\cite{Galiautdinov2012,Matsuzaki2012,Khezri2015}. On the other hand, if the qubit is initialized in the bare state $\ket{1g}$, the qubit time evolution would be characterized by fast small-amplitude beatings between the eigenmodes, which would affect qubit gate errors~\cite{Galiautdinov2012}. According to our numerical simulations, such types of errors can be mitigated by setting the gate duration to the optimal value $t_\text{g}^\text{opt} = C_\text{g} \times 2 \pi / |\Delta| $, where the parameter $C_\text{g}$ depends on details of a particular qubit gate implementation and should be determined numerically~(\cref{appendix-fidelity}). The significance of the reported method of TLS defect detection is that it allows one to extract detailed information about off-resonant TLS defects in a given qubit which can be then used for numerical optimization of relevant gate parameters.

According to numerical simulations, similar phenomena of dynamical coupling between a qubit and an off-resonant TLS defect can be observed  when the qubit is strongly driven by other pulse sequences, such as a Rabi drive~(\cref{appendix-Rabi}).

The described approach for detection of high-frequency TLS defects complements techniques for probing low-frequency TLS defects using a spin-locking pulse sequence~\cite{Yan2013}. For low-frequency TLS signatures, the condition of resonant qubit-defect interaction would be given by the Hartmann-Hahn-type equation~$\Omega = \omega_\text{TLS}$, and, therefore, in contrast to high-frequency TLS defects, positions of low-frequency TLS signatures would not change significantly in the narrow range of magnetic flux biases used in this work.

\section{Conclusions}

We introduced and experimentally demonstrated a method of high-frequency TLS defect spectroscopy in superconducting qubits that allowed us to distinguish between defects with different types of qubit-defect interaction. Using this method, we succeeded in the unambiguous detection of critical-current-fluctuation TLS defects that remained elusive until now.
The described technique should be also suitable for detection of other types of high-frequency defects, such as spin defects~\cite{Saito2013,Bienfait2016,Toida2019,Ranjan2020,OSullivan2020}.
We envision that the reported method will become a standard protocol for systematic studies of high-frequency defects in both flux-tunable and fixed-frequency superconducting qubits, revealing new insights into the microscopic origin of TLS defects and their mitigation strategies. The presented approach complements methods for the characterization of other types of noises in superconducting qubits, facilitating further improvement in the performance of superconducting quantum processors.

\begin{acknowledgments}

We appreciate William J. Munro for helpful discussions and granting the access to an HPC server for numerical simulations. We thank Aijiro Saito for his technical support with the qubit fabrication. Y.M. acknowledges the support by Leading Initiative for Excellent Young Researchers MEXT Japan and JST PRESTO
(Grant No. JPMJPR1919) Japan. This work was partially supported by JST CREST (JPMJCR1774) and JST Moonshot R\&D (JPMJMS2067).

\end{acknowledgments}

\appendix

\section{TLS defect Hamiltonian\label{appendix-defect-Hamiltonian}}

In the standard tunneling model, a two-level-system defect can occupy one of the two position states $\ket{L}$ and $\ket{R}$ corresponding to the minima of the double-well potential with the tunneling rate $\Delta_0$ and assymetry energy $\hbar \varepsilon$~\cite{Muller2019}. The effective Hamiltonian in the position basis is given by
\begin{equation}
\hat{H}_\text{TLS} = \frac{1}{2} \hbar \varepsilon \hat{\sigma}_z^{(p)} + \frac{1}{2} \hbar \Delta_0 \hat{\sigma}_x^{(p)},
\end{equation}
where $\hat{\sigma}_z^{(p)}=\ket{R}\bra{R}-\ket{L}\bra{L}$, and $\hat{\sigma}_x^{(p)} = \ket{R}\bra{L}+\ket{L}\bra{R}$. 

The transformation from the position basis to the eigenstate basis is performed by the rotation by the angle $\theta = \arctan(\Delta_0/\varepsilon)$:
\begin{equation}
\begin{split}
\hat{\sigma}_x^{(p)} & = \cos \theta \hat{\sigma}_x + \sin \theta \hat{\sigma}_z, \\
\hat{\sigma}_y^{(p)} & =\hat{\sigma}_y, \\
\hat{\sigma}_z^{(p)} & = \cos \theta \hat{\sigma}_z - \sin \theta \hat{\sigma}_x,
\end{split}
\end{equation} 
and the Hamiltonian of the TLS defect in the eigenstate basis is given by
\begin{equation}
\hat{H}_\text{TLS} = \frac{1}{2} \hbar \sqrt{\varepsilon^2 + \Delta_0^2} \hat{\sigma}_z = \frac{1}{2} \hbar \omega_\text{TLS} \hat{\sigma}_z.
\end{equation}

For simplicity, the assymetry parameter $\varepsilon$ is assumed to be negligible throughout this work, $\varepsilon \approx 0$, and, hence, $\theta = \pi/2$. Then the relations between operators in position and eigenstate bases are given by
\begin{equation}
\hat{\sigma}_x^{(p)} = \hat{\sigma}_z, \; \hat{\sigma}_y^{(p)} = \hat{\sigma}_y, \; \hat{\sigma}_z^{(p)} = - \hat{\sigma}_x.
\label{position-to-eigenstate}
\end{equation}

\begin{figure}[t!]
    \centering
    \includegraphics[width=\columnwidth]{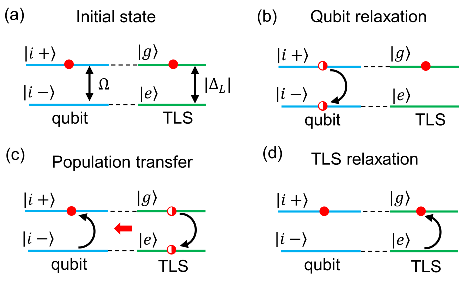}
    \caption{The population transfer mechanism in the case of a ``hot'' TLS defect ($\Delta_\text{L}<0$). Energy level diagrams in the rotating frame are shown at different stages of the population transfer process. (a) The qubit and defect are initialized to the states $\ket{i+}$ and $\ket{g}$, respectively. (b) The qubit population is distributed between the state $\ket{i+}$ and the state $\ket{i-}$ due to the qubit relaxation process. (c) The population is transferred from the TLS defect to the qubit. (d) The TLS defect is reset to the state $\ket{g}$ by the defect energy relaxation process.}
    \label{fig-hot-defect}
\end{figure}

\section{Rotating wave approximation\label{RWA}}

The transformation from the laboratory frame Hamiltonian $\hat{H}$ to the rotating frame Hamiltonian $\hat{H}^\text{R}$ is determined by 
\begin{equation}
\hat{H}^\text{R}  = \hat{U} \hat{H} \hat{U}^\dagger - \hat{R},
\end{equation}
where the unitary transformation $\hat{U}$ is given by
\begin{equation}
\hat{U} = e^{i\hat{R}t/\hbar}.
\end{equation}

In the case of the charge-fluctuation TLS defect, we use the following operator:
\begin{equation}
\hat{R}_\text{C}=\hbar \omega_\text{q} \left(\hat{b}^\dagger \hat{b} + \frac{1}{2}\hat{\sigma}_z\right).
\label{operator-charge}
\end{equation}

It can be shown that the $\hat{R_\text{C}}$ operator results in the following RWA transformation:
\begin{equation}
\begin{split}
\hat{U} \hat{b} \hat{U}^\dagger & = \hat{b} e^{-i\omega_\text{q} t}, \\
\hat{U}  \hat{\sigma}_- \hat{U}^\dagger & = \hat{\sigma}_- e^{-i\omega_\text{q} t}.
\end{split}
\end{equation}

In the case of the critical-current-fluctuation TLS defect, we use the following operator:
\begin{equation}
\hat{R}_\text{I}=\hbar \omega_\text{q} (\hat{b}^\dagger \hat{b} + \hat{\sigma}_z),
\label{operator-critical-current}
\end{equation}
and the corresponding RWA transformation is given by
\begin{equation}
\begin{split}
\hat{U} \hat{b} \hat{U}^\dagger& = \hat{b} e^{-i\omega_\text{q} t}, \\
\hat{U} \hat{\sigma}_- \hat{U}^\dagger & = \hat{\sigma}_- e^{-2i\omega_\text{q} t}.
\end{split}
\end{equation}

\begin{figure}[t!]
    \centering
    \includegraphics[width=\columnwidth]{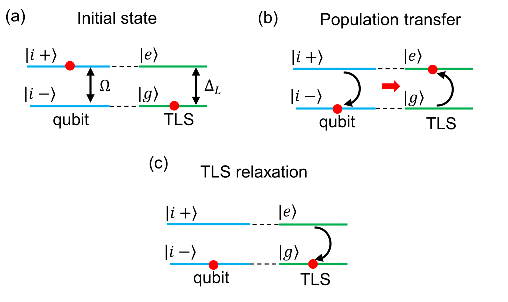}
    \caption{The population transfer mechanism in the case of a ``cold'' TLS defect ($\Delta_\text{L}>0$). Energy level diagrams in the rotating frame are shown at different stages of the population transfer process. (a) The qubit and defect are initialized to the states $\ket{i+}$ and $\ket{g}$, respectively. (b) The population is transferred from the qubit to the TLS defect. (c) The TLS defect is reset to the state $\ket{g}$ by the defect energy relaxation process.}
    \label{fig-cold-defect}
\end{figure}

\section{Population transfer in the case of ``hot'' and ``cold'' TLS defects\label{hot-defect}}

Figure~\ref{fig-hot-defect} illustrates the population transfer mechanism in the rotating frame in the case of a ``hot'' TLS defect. Here, we consider a charge-fluctuation TLS defect with the negative detuning, $\Delta_\text{L}<0$. As described in the main text, it is assumed that there is no dephasing, and the collapse operators for qubit and defect relaxation processes are chosen in the form of $\hat{C}_1 \propto \hat{b}$ and $\hat{C}_2 \propto\hat{\sigma}_-$, respectively. Initially, in the rotating frame, the excited states of the qubit and defect are populated~[\cref{fig-hot-defect}(a)], and, hence, there is no population transfer between them. Here, the excited state of the TLS defect in the rotating frame $\ket{g}$ corresponds to the defect ground state in the laboratory frame, and, therefore, the TLS defect in the state $\ket{g}$ does not relax. After the qubit relaxes from $\ket{i+}$ to a mixed state~[\cref{fig-hot-defect}(b)], the population transfer between the qubit and TLS defect occurs: the qubit and defect are flipped to the states $\ket{i+}$ and $\ket{e}$, respectively~[\cref{fig-hot-defect}(c)]. Finally, the TLS defect relaxes to the state $\ket{g}$~[\cref{fig-hot-defect}(d)]. The described mechanism explains the high level of the population of the state $\ket{i+}$ observed in the case of $\Delta_\text{L}<0$. The detailed description of that process would require solving rate equations, which is beyond the scope of this work.

For comparison,~\cref{fig-cold-defect} demonstrates the population transfer mechanism in the rotating frame in the case of a ``cold'' TLS defect ($\Delta_\text{L}>0$).

\section{Effect of counter-rotating couplings\label{counter-rotating-coupling}}

In this section, we consider a different model of the coupling between a qubit and a critical-current defect by taking into account counter-rotating terms omitted in~~\cref{Hamiltonian-current}. We start from the Hamiltonian given by~\cref{Hamiltonian-current-full}. The nonlinear interaction term given by~\cref{current-defect} can be expanded in the following form:
\begin{equation}
\hat{H}_\text{int}^{\text{(I)}} = \hbar g_I^{(2)} \hat{\sigma}_x \left( \left(\hat{b}^\dagger\right)^2 + \hat{b}^2 + 2\hat{b}^\dagger\hat{b} +1 \right).
\end{equation}

In the main text, the term proportional to $\hat{b}^\dagger\hat{b}$ was eliminated by using the RWA. In this section, we take into account the effect of this interaction beyond the RWA. We consider the following interaction Hamiltonian:  
\begin{equation}
\hat{H}_\text{int}^\prime = 2 \hbar g_I^{(2)} \hat{\sigma}_x \hat{b}^\dagger\hat{b}.
\end{equation}

By truncating the qubit Hilbert space to the lowest two states, we can use the following relations between bosonic operators and qubit Pauli operators $\hat{\tau}_{x,y,z}$ in the eigenstate basis $\{\ket{0},\ket{1}\}$:
\begin{equation}
\begin{split}
& \hat{b}  = \hat{\tau}_+ = \frac{1}{2} (\hat{\tau}_x+i\hat{\tau}_y), \\
& \hat{b}^\dagger  = \hat{\tau}_- = \frac{1}{2} (\hat{\tau}_x-i\hat{\tau}_y), \\
& \hat{b}^\dagger \hat{b} = \frac{1}{2}(1-\hat{\tau}_z). \\
\end{split}
\end{equation}

The interaction term between the qubit and a critical-current defect can be written as:
\begin{equation}
\hat{H}_\text{int}^\prime = -\hbar g_I^{(2)} \hat{\tau}_z\hat{\sigma}_x.
\end{equation}

Then, the system Hamiltonian is given by
\begin{equation}
\begin{split}
\hat{H} & = -\frac{1}{2}\hbar \omega_\text{q} \hat{\tau}_z + \frac{1}{2} \hbar \omega_\text{TLS} \hat{\sigma}_z \\ 
& -\hbar g_I^{(2)} \hat{\tau}_z\hat{\sigma}_x + \hbar \Omega \cos (\omega_\text{q}t) \hat{\tau}_y.
\end{split}
\end{equation}

In the rotating frame defined by the following rotation operator
\begin{equation}
\hat{R} =-\frac{1}{2}\hbar \omega_\text{q} \hat{\tau}_z + \frac{1}{2} \hbar \omega_\text{TLS} \hat{\sigma}_z,
\end{equation}
the system Hamiltonian is given by
\begin{equation}
\begin{split}
& \hat{H}^{(1)} = -\hbar g_I^{(2)} \hat{\tau}_z \left( \hat{\sigma}_+ e^{i \omega_\text{TLS}t} + \hat{\sigma}_- e^{-i \omega_\text{TLS}t}\right)+\frac{\hbar\Omega}{2}\hat{\tau}_y \\
& + \frac{i \hbar \Omega}{4} \left((\hat{\tau}_x - i\hat{\tau}_y)e^{+2 i \omega_\text{q}t} - (\hat{\tau}_x+i\hat{\tau}_y) e^{-2 i \omega_\text{q}t} \right).
\end{split}
\end{equation}

By changing the qubit basis from the $\hat{\tau}_z$ basis $\{\ket{0},\ket{1}\}$ to the $\hat{\tau}_y$ basis $\{\ket{i-},\ket{i+}\}$, we transform the operators $\{\hat{\tau}_x, \hat{\tau}_y, \hat{\tau}_z\}$ to $\{-\hat{s}_y, -\hat{s}_z, \hat{s}_x\}$ and obtain
\begin{equation}
\begin{split}
& \hat{H}^{(1)} = -\hbar g_I^{(2)} \hat{s}_x \left( \hat{\sigma}_+ e^{i \omega_\text{TLS}t} + \hat{\sigma}_- e^{-i \omega_\text{TLS}t}\right)-\frac{\hbar\Omega}{2} \hat{s}_z \\
& - \frac{i \hbar \Omega}{4} \left( (\hat{s}_y -i \hat{s}_z) e^{2 i \omega_\text{q}t} - (\hat{s}_y+i \hat{s}_z )e^{-2 i \omega_\text{q}t} \right).
\end{split}
\end{equation}

By going to another rotation frame defined by the operator given by
\begin{equation}
\hat{R'} = -\frac{\hbar\Omega}{2} \hat{s}_z,
\end{equation}
we obtain 
\begin{equation}
\begin{split}
\hat{H}' & = -\hbar g_I^{(2)} (\hat{s}_+ e^{-i \Omega t} + \hat{s}_- e^{i \Omega t}) \left( \hat{\sigma}_+ e^{i \omega_\text{TLS}t} + \hat{\sigma}_- e^{-i \omega_\text{TLS}t}\right) \\
& + \frac{\hbar \Omega}{4} \left((\hat{s}_- e^{i \Omega t} - \hat{s}_+ e^{-i \Omega t}) e^{2 i \omega_\text{q}t} - \hat{s}_z e^{2 i \omega_\text{q}t} + h.c.\right).
\end{split}
\end{equation}

After some rearrangement, the Hamiltonian can be written in the form
\begin{equation}
\begin{split}
\hat{H}'(t) & = -\hbar g_I^{(2)} (\hat{s}_+ \hat{\sigma}_+ e^{i (\omega_\text{TLS} - \Omega)t} + \hat{s}_- \hat{\sigma}_+ e^{i (\omega_\text{TLS}+\Omega)t} + h.c.) \\
& + \frac{\hbar \Omega}{4} (  \hat{s}_- e^{i (2\omega_\text{q} + \Omega) t} - \hat{s}_+ e^{i (2\omega_\text{q}-\Omega) t}- \hat{s}_z e^{2 i \omega_\text{q}t} + h.c.).
\end{split}
\end{equation}

To simplify the calculation, we will assume
\begin{equation}
\Omega = \omega_\text{TLS} - 2\omega_\text{q}.
\label{cond1}
\end{equation}

Then, the Hamiltonian can be written as
\begin{equation}
\begin{split}
\hat{H}'(t) & = -\hbar g_I^{(2)} (\hat{s}_+ \hat{\sigma}_+ e^{2 i \omega_\text{q} t} + \hat{s}_- \hat{\sigma}_- e^{-2 i \omega_\text{q} t}) \\
& - \frac{\hbar \Omega}{4} (\hat{s}_z e^{2 i \omega_\text{q}t} + \hat{s}_z e^{-2 i \omega_\text{q}t} ) + \dotso ,
\end{split}
\label{hamiltonian-counter-rotating-simplified}
\end{equation}
where we omitted irrelevant terms.

The equation of motion is given by
\begin{equation}
i \hbar \frac{d}{dt} \ket{\psi(t)} = \hat{H}' \ket{\psi(t)}.
\end{equation}

By integration from 0 to $t$, we obtain:
\begin{equation}
\ket{\psi(t)} = \ket{\psi(0)} - \frac{i}{\hbar} \int_0^t \hat{H}'(t') \ket{\psi(t')}dt'.
\end{equation}

By assuming that $H'$ is small, we use an iteration procedure to write the solution in the form:
\begin{equation}
\begin{split}
\ket{\psi(t)} & = \ket{\psi(0)} - \frac{i}{\hbar} \int_0^t \hat{H}'(t') \ket{\psi(0)}dt' \\
&-\frac{1}{\hbar^2}\int_0^t \hat{H}'(t') dt'\int_0^{t'}  \hat{H}'(t'') \ket{\psi(0)}dt''.
\end{split}
\label{psi}
\end{equation}

Since $\hat{H}'$ contains only fast oscillating terms, we neglect the second term in the right-hand side of~\cref{psi}:
\begin{equation}
\frac{i}{\hbar} \int_0^t \hat{H}'(t') \ket{\psi(0)}dt' \approx 0.
\end{equation}

\begin{figure*}[t!]
    \centering
    \includegraphics[width=.6\textwidth]{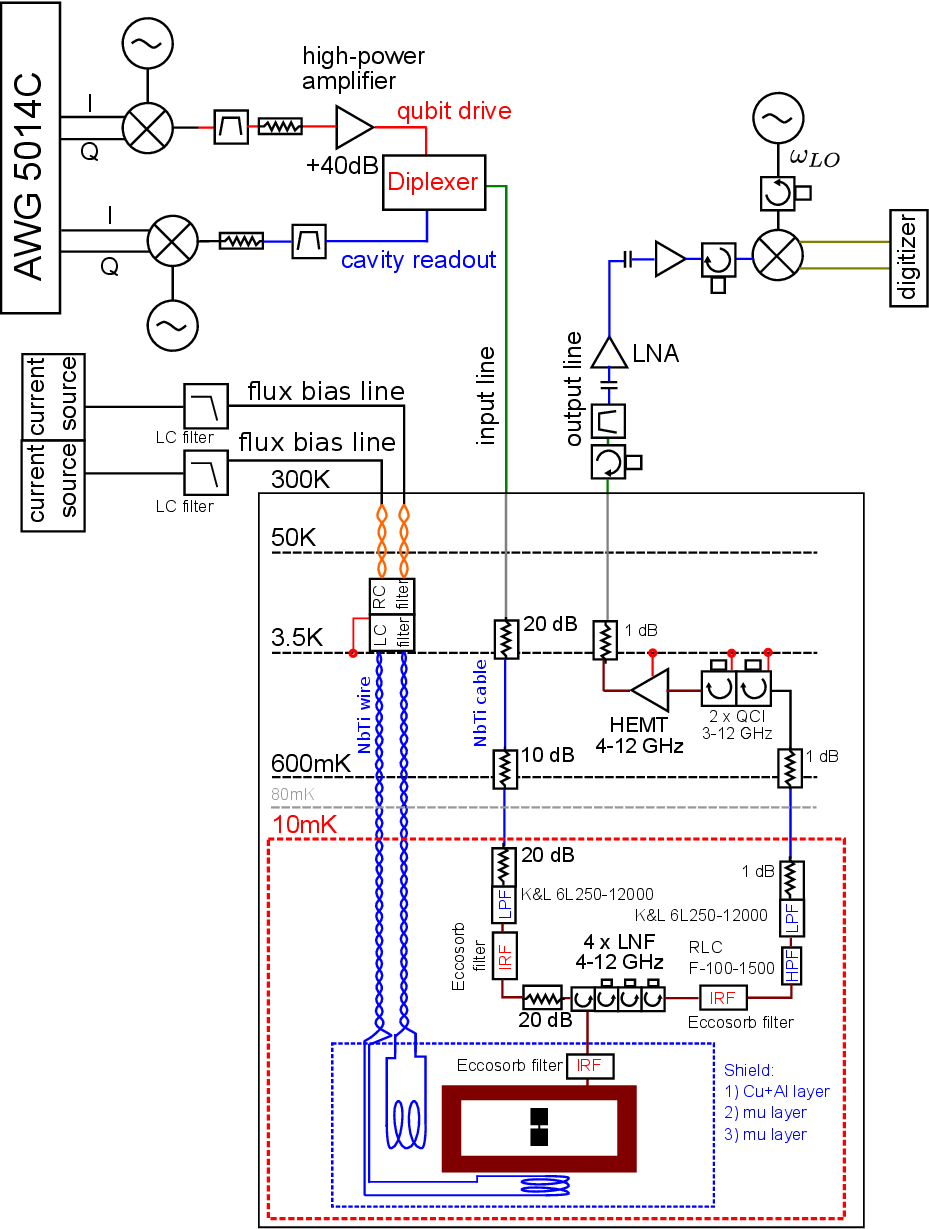}
    \caption{The schematic of the measurement setup.}
    \label{fig-measurement-setup}
\end{figure*}

Using~\cref{hamiltonian-counter-rotating-simplified}, we write the last term in the right-hand side of~\cref{psi} in the form:
\begin{equation}
\begin{split}
& -\frac{1}{\hbar^2}\int_0^t \hat{H}'(t') dt'\int_0^{t'}  \hat{H}'(t'') \ket{\psi(0)}dt'' \\
& \approx -\frac{i}{\hbar}  \int_0^t \hat{H}_\text{eff} \ket{\psi(0)} dt',
\end{split}
\end{equation}
where we drop fast oscillating terms. Here, the effective Hamiltonian is given by
\begin{equation}
\hat{H}_\text{eff} = -\hbar \frac{g_I^{(2)} \Omega}{4\omega_\text{q}}( \hat{s}_- \hat{\sigma}_- + \hat{s}_+ \hat{\sigma}_+  ).
\end{equation}

Repeating the same procedure for the condition
\begin{equation}
\Omega = 2\omega_\text{q}-\omega_\text{TLS},
\label{cond2}
\end{equation}
we obtain the effective Hamiltonian
\begin{equation}
\hat{H}_\text{eff} = \hbar \frac{g_I^{(2)} \Omega}{4\omega_\text{q}}( \hat{s}_- \hat{\sigma}_+ + \hat{s}_+ \hat{\sigma}_- ).
\end{equation}

Thus, if the condition~\cref{cond1} or~\cref{cond2} is met, there is an effective coupling between the qubit and defect due to counter-rotating terms, but the effective coupling strength is given by 
\begin{equation}
g_\text{eff} =  \frac{g_I^{(2)} \Omega}{4\omega_\text{q}},
\end{equation}
which is smaller than the coupling strength value given by~\cref{effective-coupling-strength}.

\begin{figure}[t!]
    \centering
    \includegraphics{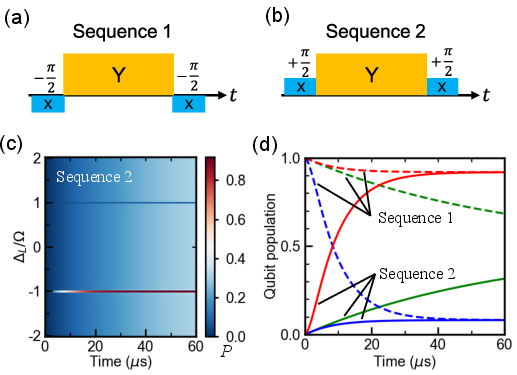}
    \caption{(a) The schematic of the standard spin-locking sequence used in measurements reported in the main text (Sequence 1). (b) The schematic of the modified spin-locking sequence (Sequence 2). (c) Results of numerical simulations of the system evolution when the qubit is driven by the Sequence 2. The population of the qubit state $\ket{i+}$ is shown. The initial qubit state is $\ket{i-}$. The other parameters are the same as used for the results shown in~\cref{fig1}(c,d). (d) The comparison of the qubit response at the qubit-defect detunings $\Delta_L = \Omega$ (blue lines), $\Delta_L = -\Omega$ (red lines), and $\Delta_L \neq \pm \Omega$ (green lines). Solid lines correspond to the Sequence 2, dashed lines correspond to the Sequence 1.}
    \label{phase-cycling-theory}
\end{figure}

\section{Measurement setup\label{appendix-measurement-setup}}
The measurement setup was similar to the one used in the previous works~\cite{Abdurakhimov2019,Abdurakhimov2020}, with a few modifications made in the measurement circuit (\cref{fig-measurement-setup}). A broadband 2 -- 18 GHz IQ mixer Marki Microwave MMIQ-0218L was used to generate the qubit drive. A microwave diplexer Marki Microwave DPX-0508 was used to combine the qubit drive and cavity readout signals. The qubit drive was applied via the low-pass port of the diplexer with the pass band of DC -- 5 GHz, while the cavity readout was applied through the high-pass port of the diplexer with the pass band of 8 -- 18 GHz.

\section{Additional experimental results}

\subsection{Phase-cycling procedure\label{phase-cycling}}

Besides the standard spin-locking sequence (Sequence 1) described in the main text and shown in~\cref{fig3}(d) and~\cref{phase-cycling-theory}(a), we also performed measurements using a modified spin-locking sequence (Sequence 2) where the phases of X-pulses were inverted~[\cref{phase-cycling-theory}(b)]. In the case of the modified spin-locking pulse sequence, the first X-pulse rotates the qubit state vector around the X-axis by the angle $\pi/2$, leaving the qubit in the state $\ket{i-}$. We performed numerical simulations of the evolution of the qubit state under a strong Y-pulse drive for the case when the qubit was coupled to a charge-fluctuation defect~[\cref{phase-cycling-theory}(c)]. Except for the initial qubit state, all other simulation parameters were the same as used for the results shown in~\cref{fig1}(c,d). It was found that, although the initial states were different, the stationary populations of $\ket{i+}$ state for the Sequence 1 and Sequence 2 were the same for sufficiently long Y-pulse durations~[\cref{phase-cycling-theory}(d)]:
\begin{equation}
P_+^{(S1)} = P_+^{(S2)} \quad \textrm{for} \quad t \to \infty.
\label{population1}
\end{equation}

However, the populations of the final state $\ket{1}$ are different for the Sequence 1 and Sequence 2 (after the X-pulse rotation by $-\pi/2$ and $\pi/2$, respectively):
\begin{equation}
P_1^{(S1)} = P_+^{(S1)} \quad \textrm{and} \quad P_1^{(S2)} = 1 - P_+^{(S2)}.
\end{equation}
Thus, for equal levels of qubit populations in the rotating frame given by~\cref{population1}, the final output signals of Sequence 1 and Sequence 2 have inverted ``polarities''.

Due to a possible apparatus noise, the actual population values measured in experiments can have some additional offsets:
\begin{equation}
\begin{split}
P_1^{(S1)} & = P_+^{(S1)} + N,\\
P_1^{(S2)} & = 1 - P_+^{(S2)} + N,
\label{population2}
\end{split}
\end{equation}
where $N$ is the offset due to the apparatus noise.

It follows from~\cref{population1,population2} that the apparatus noise can be eliminated by calculating the pure population value
\begin{equation}
P_1 = \frac{1}{2} + \frac{1}{2} (P_1^{(S1)} - P_1^{(S2)}).
\end{equation}
The data shown in~\cref{fig4}(a,b) was processed using the described procedure.

\subsection{Rabi frequency calibration\label{calibration}}
\begin{figure}[h!]
    \centering
    \includegraphics{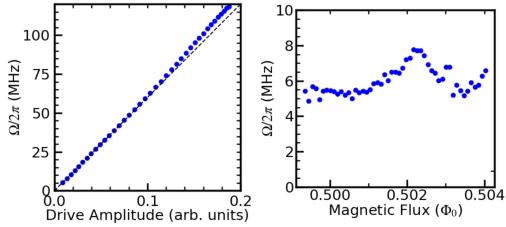}
    \caption{(a) The dependence of the Rabi frequency on the qubit drive amplitude at the optimal point $\Phi\approx 0.5 \Phi_0$. The black dashed line corresponds to a linear function plot,  added for eye guidance. (b) The dependence of the Rabi frequency on the applied magnetic flux for a fixed qubit drive amplitude.}
    \label{fig-rabi-calibration}
\end{figure}

The Rabi frequency $\Omega$ was calibrated in separate experiments by measuring the period of Rabi oscillations of the qubit. Dependencies of the Rabi frequency on the drive amplitude and applied magnetic flux are shown in~\cref{fig-rabi-calibration}(a) and~\cref{fig-rabi-calibration}(b), respectively. The slight deviation of the Rabi frequency from a linear fit at high drive amplitudes was due to the low fitting accuracy which was caused by the low sampling rate of the data.

\begin{figure}[h!]
    \centering
    \includegraphics[width=\columnwidth]{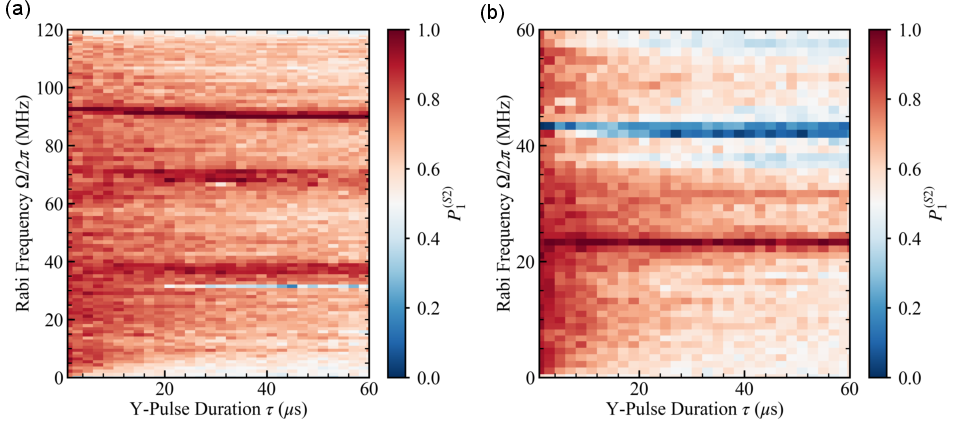}
    \caption{Typical results of spin-locking measurements using the modified pulse-sequence (Sequence 2) (a) at the optimal flux bias point $0.5\Phi_0$, and (b) at the flux bias $0.5039\Phi_0$. The qubit was initialized in the $\ket{i-}$ state in the rotating frame.}
    \label{experiment-raw-data}
\end{figure}

\begin{figure}[h!]
    \centering
    \includegraphics[width=\columnwidth]{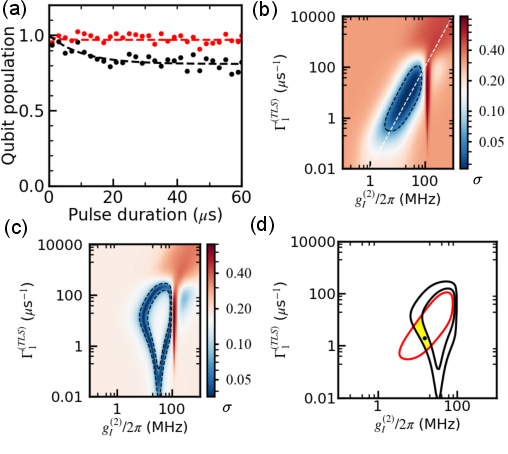}
    \caption{Results of the fitting of time-domain spin-locking signals corresponding to a critical-current-fluctuation TLS defect. The experimental data points were extracted from the results shown in \cref{experiment-raw-data}(b). The qubit was initialized in the $\ket{i-}$ state in the rotating frame. The magnetic flux bias was $0.5039 \Phi_0$. At the given bias, the detuning between the qubit and critical-current-fluctuation TLS defect was $\Delta_\text{NL}/2\pi\approx$\,23.4\,MHz. (a) The qubit population decay at the Rabi frequency $\Omega/2\pi \approx$\,23.4\,MHz (red dots) and $\Omega/2\pi \approx$\,22.2\,MHz (black dots) where the effective qubit-defect detuning was $\delta \Omega/2\pi =$\,0\,MHz and $\delta \Omega/2\pi =$\,1.2\,MHz, respectively. Dashed lines represent results of numerical simulations described in the text. (b) The RMS fitting deviation (residual) $\sigma$ as a function of the fitting parameters $g_I^{(2)}$ and $\Gamma_{1\text{TLS}}$ for the experimental data obtained at $\delta \Omega/2\pi =$\,0\,MHz. The dashed white line corresponds to the scaling law $\mathcal{F}\propto (g_I^{(2)})^2/ \Gamma_{1\text{TLS}}$ (added for eye guidance). The dashed black curve bounds the region of optimal fitting parameters. (c) The RMS fitting deviation $\sigma$ for the experimental data obtained at $\delta \Omega/2\pi = $\,1.2\,MHz. The dashed black curves bound the regions of optimal fitting parameters. (d) The regions of optimal fitting parameters for $\delta \Omega/2\pi =$\,0\,MHz (red curve) and $\delta \Omega/2\pi = $\,1.2\,MHz (black curve), respectively. The yellow area corresponds to the overlap between the regions. The black dot corresponds to the fitting values $g_I^{(2)}$ and $\Gamma_{1\text{TLS}}$ used to plot fitting curves shown in~\cref{fitting}(a).} 
    \label{fitting}
\end{figure}

\begin{figure}[h!]
    \centering
    \includegraphics[width=0.9\columnwidth]{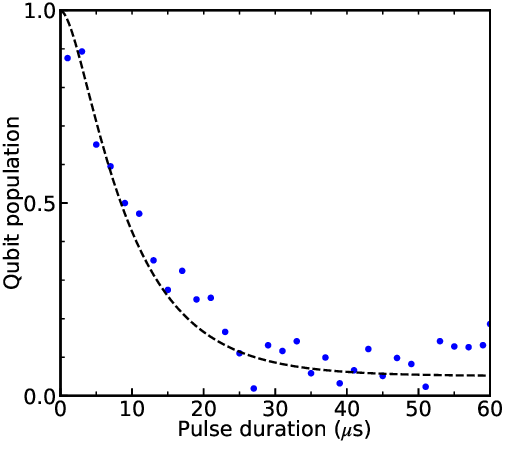}
    \caption{Results of the fitting of the time-domain spin-locking signal corresponding to a charge-fluctuation TLS defect. The experimental data points were extracted from the results shown in \cref{experiment-raw-data}(b). The qubit was initialized in the $\ket{i-}$ state in the rotating frame. The magnetic flux bias was $0.5039 \Phi_0$.}
    \label{charge-defect-fitting}
\end{figure}

\subsection{Time-domain spin-locking measurements\label{preliminary}}

Time-domain spin-locking measurements were performed using the modified spin-locking sequence (Sequence 2) described in~\cref{phase-cycling}. Typical results are shown in~\cref{experiment-raw-data}. Pronounced horizontal spectral line were observed at Rabi frequencies where the qubit was dynamically coupled to TLS defects. The Y-pulse duration 60\,$\mu$s was enough to reach the stationary levels of the qubit population.

The parameters of the qubit-defect coupled system can be estimated by fitting the experimental data using the theoretical models described in the main text. 
Figure~\ref{fitting} shows results of the fitting of time-domain spin-locking data corresponding to a critical-current fluctuation TLS defect. 
The qubit-defect coupling strength $g_I^{(2)}$ and defect relaxation rate $\Gamma_{1\text{TLS}}$ were used as fitting parameters, and the fitting was performed using the following procedure. 

First, by comparing the positions of the spectral line TLS5 in~\cref{fig4}(b) and spectral lines in~\cref{experiment-raw-data}(b) at the flux bias $0.5039\Phi_0$, we identified that the pronounced spectral line, which was observed in~\cref{experiment-raw-data}(b) at $\Omega/2\pi \approx$\,23.4\,MHz, corresponded to the critical-current-fluctuation TLS defect (it should be noted that~\cref{fig4}(a,b) were obtained using the standard spin-locking sequence, while~\cref{experiment-raw-data} was obtained using the modified spin-locking sequence, and, hence, the ``polarities'' of spin-locking signals were inverted). Thus, at the given flux bias, the qubit-defect detuning was $\Delta_\text{NL}/2\pi\approx$\,23.4\,MHz. 

Second, we extracted two data sets from the data shown in~\cref{experiment-raw-data}(b): the spin-locking signal at $\Omega/2\pi\approx$\,23.4\,MHz and  another one at $\Omega/2\pi \approx$\,22.2\,MHz which corresponded to effective qubit-defect detunings $\delta \Omega/2\pi =$\,0\,MHz and $\delta \Omega/2\pi =$\,1.2\,MHz, respectively [\cref{fitting}(a)]. Here, the effective qubit-defect detuning in the rotating frame is defined as $\delta \Omega = \Delta_\text{NL} - \Omega$.

Third, we calculated the root-mean-square (RMS) fitting deviation (fitting residual) $\sigma$ for each data set using the following expression:
\begin{equation}
\sigma = \sqrt{\langle \left[ Y_\text{exp} - Y_\text{fit} (g_I^{(2)},\Gamma_{1\text{TLS}} ) \right]^2 \rangle},
\end{equation}  
where $Y_\text{exp}$ corresponds to a given data set shown in~\cref{fitting}(a), and $Y_\text{fit}$ is the result of numerical simulations for given fitting parameters $g_I^{(2)}$ and $\Gamma_{1\text{TLS}}$ [\cref{fitting}(b,c)]. To calculate $Y_\text{fit}$, we numerically solved the master equation for the Hamiltonian given by~\cref{Hamiltonian-current} using the following parameters: $\Delta_\text{NL}/2\pi = $\,23.4\,MHz, $A/2\pi =$\,1\,GHz, $\Gamma_{1\text{q}} = 1/T_1 \approx $\,0.02\,$\mu$s$^{-1}$, and $\Gamma_{2\text{q}} = 1/T_{2\text{E}} \approx $\,0.03\,$\mu$s$^{-1}$. The qubit states were truncated to the lowest three levels. The collapse operators were $\hat{C}_1 = \sqrt{\Gamma_{1\text{q}}} \hat{b}$, $\hat{C}_2 = \sqrt{2\Gamma_{2\text{q}}} \hat{b}^\dagger \hat{b}$, and $\hat{C}_3=\sqrt{\Gamma_{1\text{TLS}}}\hat{\sigma}_-$. Here, we phenomenologically introduced the pure dephasing term $\hat{C}_2$ to improve the fitting accuracy for the data at $\delta \Omega/2\pi =$\,1.2\,MHz. The shape of the pure dephasing term was chosen to be similar to the term typically used for a two-level qubit: $\hat{C}_2 = \sqrt{\Gamma_{2\text{q}}/2} \hat{\tau}_z \approx \sqrt{2\Gamma_{2\text{q}}} \hat{b}^\dagger \hat{b}$. The qubit was initialized in its ground state $\ket{i-}$ in the rotating frame, the defect was initialized in its ground state $\ket{g}$, and the population of the qubit state $\ket{i-}$ was calculated.
Figures~\ref{fitting}(b) and~\ref{fitting}(c) shows fitting deviations for the data obtained at $\delta \Omega/2\pi =$\,0\,MHz and $\delta \Omega/2\pi =$\,1.2\,MHz, respectively. The regions of optimal fitting parameters were determined by plotting contour plots at the level of $2 \sigma_\text{min}$, where $\sigma_\text{min}$ was the minimum value of the fitting deviation for a given data set. In the case of $\delta \Omega/2\pi =$\,0\,MHz, the minimum fitting residual was achieved in the range of fitting parameters which follow the scaling law $\mathcal{F} \propto \left(g_I^{(2)}\right)^2 / \Gamma_{1\text{TLS}}$. This scaling was due to the fact that the qubit-defect system was in the Purcell regime with the effective qubit relaxation rate given by $ \Gamma_\text{P} = 4 g_\text{eff}^2 / \Gamma_{1\text{TLS}}$.

Finally, the region of ``global'' fitting parameters was determined by finding the overlap of the regions of ``local'' optimal fitting parameters for  $\delta \Omega/2\pi =$\,0\,MHz and $\delta \Omega/2\pi =$\,1.2\,MHz [\cref{fitting}(d)]. Then, the fitting curves shown in~\cref{fitting}(a) were calculated using the fitting parameters from that region: $g_I^{(2)}/2\pi=$\,15\,MHz, and $\Gamma_{1\text{TLS}}=$\,2\,$\mu$s$^{-1}$. 

A similar approach was used to extract characteristics of charge-fluctuation TLS defects. By fitting the time-domain spin-locking data (\cref{charge-defect-fitting}), the coupling strength and the relaxation rate for charge-fluctuation TLS defects were estimated to be $g_\text{C}/2\pi\approx$\,50\,kHz and $\Gamma_{1\text{TLS}}\approx $\,1\,$\mu$s$^{-1}$, respectively.

\begin{figure}[h!]
    \centering
    \includegraphics[width=\columnwidth]{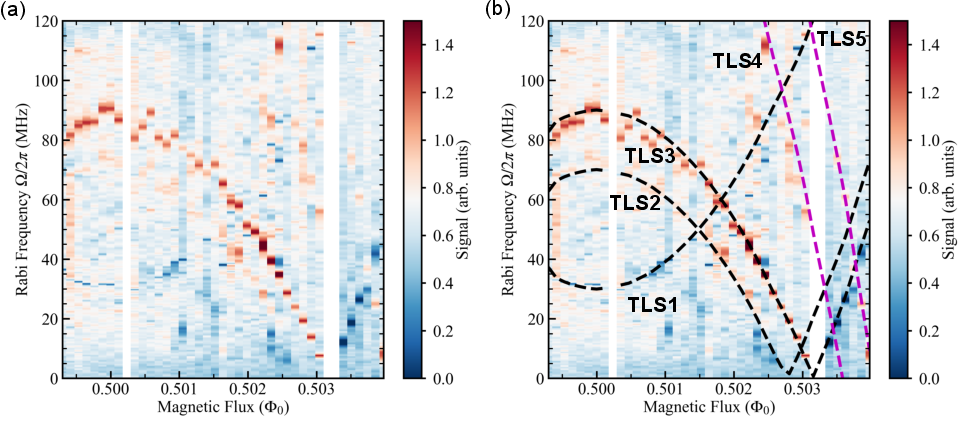}
    \caption{Results of TLS defect spectroscopy in experiments with the applied in-plane magnetic field of 0.2\,mT. The qubit was initialized in the $\ket{i-}$ state in the rotating frame. (a) The excited-state qubit population as a function of the applied magnetic flux and the qubit drive amplitude in the units of Rabi frequency. (b) The comparison of the experimental data obtained at 0.2\,mT with the positions of spectral lines TLS1--TLS5 determined in the experiments at 0\,mT. Dashed lines were plotted using the same equations and parameters as the dashed lines shown in~\cref{fig4}(b).}
    \label{inplane}
\end{figure}

\subsection{Dependence on the applied in-plane magnetic field\label{magnetic-field}}
Figure~\ref{inplane} shows results of TLS defect spectroscopy in experiments with an external magnetic field applied parallel to the qubit surface. The strength of the magnetic field was about 0.2\,mT. Experiments were performed using the modified pulse sequence described in~\cref{phase-cycling}. The positions of spectral signatures of TLS defects were roughly the same in experiments with (\cref{inplane}) and without (\cref{fig4}) the in-plane magnetic field.

\section{Additional numerical results}

\subsection{Numerical simulations of the system evolution without using RWA\label{time-depend}}
\begin{figure}[h!]
    \centering
    \includegraphics[width=\columnwidth]{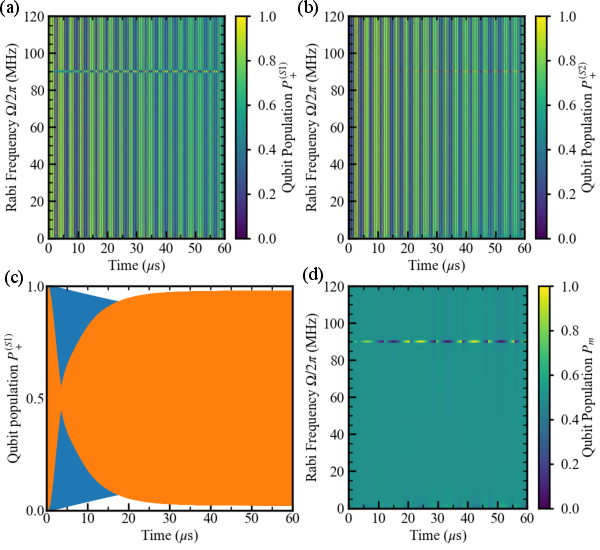}
    \caption{Time-domain results of numerical simulations of the time-dependent system Hamiltonian without using RWA. The qubit frequency was set to $\omega_\text{q}/2\pi \approx $~3.825\,GHz. Due to signal fast oscillations, \cref{time-depend-simulation}(a,b,d) include plotting artifacts in the form of low-frequency modulations. (a) Simulation results for the initial state $\ket{i+}$. The vertical stripe structure is a plotting artifact (the actual oscillation period is much shorter). (b) Simulation results for the initial state $\ket{i-}$. (c) Traces extracted from the data shown in~\cref{time-depend-simulation}(a) at the Rabi frequency $\Omega/2\pi = $\,0\,MHz (blue line) and $\Omega/2\pi = $\,90\,MHz (orange line). At $\Omega/2\pi = $\,90\,MHz, the driven qubit evolution was affected by the interaction between the qubit and TLS defect. (d) The arithmetic mean of data shown in~\cref{time-depend-simulation}(a) and~\cref{time-depend-simulation}(b). }
    \label{time-depend-simulation}
\end{figure}

As described in the main text, charge-fluctuation and critical-current-fluctuation TLS defects can be modeled separately using RWAs with two different rotation operators given by~\cref{operator-charge} and~\cref{operator-critical-current}, respectively. However, it is not possible to model defects with different types of qubit-defect interactions simultaneously in the same rotating frame. In order to numerically reproduce the experimental results which included spectral lines of both charge and critical-current TLS defects, we solved a Lindblad master equation for the following time-dependent Hamiltonian without using RWA:
\begin{equation}
\begin{split}
\frac{\hat{H}_t}{\hbar} = & \left( \omega_q \left(\Phi_e \right) - \frac{A}{2}\right) \hat{b}^\dagger \hat{b} + \frac{A}{2} \left( \hat{b}^\dagger \hat{b} \right)^2 \\
& + \frac{\omega_\text{TLS3}}{2}  \hat{\sigma}_z^{(3)} + \frac{\omega_\text{TLS5}}{2}  \hat{\sigma}_z^{(5)} \\
& + i g_C \hat{\sigma}_x^{(3)}\left( \hat{b}^\dagger - \hat{b} \right) + g_I \hat{\sigma}_x^{(5)} \left(\hat{b}^\dagger + \hat{b} \right)^2 \\
& + i \Omega \cos {(\omega_q t)} \left( \hat{b}^\dagger - \hat{b}\right).
\label{time-depend-Hamiltonian}
\end{split}
\end{equation}
Here, the model system consisted of the qubit, charge defect TLS3, and critical-current defect TLS5 (other defects were not included in the numerical model in order to minimize the calculation time). We also took into account the dependence of the qubit frequency on the applied flux bias $\Phi_e$. It should be noted that a complete derivation of a flux-dependent qubit Hamiltonian in terms of bosonic operators is beyond the scope of this paper, and the first line of the Hamiltonian given by~\cref{time-depend-Hamiltonian} represents a phenomenological model of a Duffing oscillator with a flux-dependent resonant frequency. The dependence of the qubit frequency on the applied flux bias is shown in \cref{fig3}(c) in the main text. At each value of the qubit frequency $\omega_q$, the system evolution was calculated for two initial qubit states $\ket{i+}$ and  $\ket{i-}$,  using the following parameters: $\omega_\text{TLS3}/2\pi \approx $\,3.915\,GHz, $\omega_\text{TLS5}/2\pi \approx $\,7.945\,GHz, $A/2\pi =$\,1\,GHz, $\Gamma_{1\text{q}} = 1/T_1 \approx $\,0.02\,$\mu$s$^{-1}$, $\Gamma_{1\text{TLS3}} =\Gamma_{1\text{TLS5}}=$\,1\,$\mu$s$^{-1}$, $g_C/2\pi=$\,100\,kHz, and $g_I/2\pi=$\,20\,MHz. For the initial states $\ket{i+}$ and  $\ket{i-}$, the population of the qubit state $\ket{i+}$ was calculated as a function of the Rabi frequency and evolution time, $P_+^{(S1)}(\omega_q,\Omega,t)$ and $P_+^{(S2)}(\omega_q,\Omega,t)$, respectively. Figure~\ref{time-depend-simulation}(a,b) shows $P_+^{(S1)}(\omega_q,\Omega,t)$ and $P_+^{(S2)}(\omega_q,\Omega,t)$, respectively, obtained at a given qubit frequency. Since simulations were performed in the laboratory frame, the qubit population signal oscillated with the frequency $\omega_q/2\pi$, and it was difficult to distinguish TLS spectral lines in 2D plots. However, the effect of the qubit-TLS interaction on the driven qubit population was clearly observed in individuals traces~[\cref{time-depend-simulation}(c)]. In order to remove the unwanted background, the modified qubit population signal was calculated using the following equation:
\begin{equation}
P_m(\omega_q,\Omega,t) = \frac{P_+^{(S1)}(\omega_q,\Omega,t) + P_+^{(S2)}(\omega_q,\Omega,t)}{2}.
\end{equation}

When there was no interaction between the qubit and a TLS defect, the signals $P_+^{(S1)}(\omega_q,\Omega,t)$ and $P_+^{(S2)}(\omega_q,\Omega,t)$ had opposite phases, $P_+^{(S1)}(\omega_q,\Omega,t)=1-P_+^{(S2)}(\omega_q,\Omega,t)$, and the modified qubit population signal was constant, $P_m(\omega_q,\Omega,t)=0.5$. Figure \ref{time-depend-simulation}(d) shows $P_m(\omega_q,\Omega,t)$ at a given qubit frequency, where we can clearly see a spectral line due to the qubit-defect interaction at the Rabi frequency $\Omega/2\pi=$\,90\,MHz. 

The data shown in~\cref{fig5}(b) in the main text represents amplitudes of the modified qubit population oscillations in the end of each simulation that were estimated by using the following equation:
\begin{equation}
A_p = \frac{P_{\text{min}}+P_{\text{max}}}{2},
\end{equation}
where $P_{\text{min}}(\omega_q,\Omega)$ and $P_{\text{max}}(\omega_q,\Omega)$ are the minimum and maximum values of the signal oscillations in the end of a simulation, respectively:
\begin{equation}
P_{\text{min}}(\omega_q,\Omega) = \min_{ t_1 - 2 T \leq t \leq t_1} P_m(\omega_q,\Omega,t),
\end{equation}
and
\begin{equation}
P_{\text{max}}(\omega_q,\Omega) = \max_{ t_1 - 2 T \leq t \leq t_1} P_m(\omega_q,\Omega,t),
\end{equation}
Here, $t_1$ is the simulation end time, and $T$ is the oscillation period.

\subsection{Gate errors due to the interaction between a qubit and an off-resonant TLS defect\label{appendix-fidelity}}
In this section, we describe numerical simulations of single-qubit gate errors for the system consisting of the qubit coupled to an off-resonant TLS defect. For simplicity, it is assumed that the defect is a charge-fluctuation TLS defect. We calculate the minimum gate error $p$ as a function of the gate duration $t_\text{g}$ and qubit-defect detuning $\Delta$ using the procedure described below. We assume here that the value of the qubit-defect coupling strength is $g/2\pi=$\,500\,kHz which is larger than the one observed in this work, but it is still within the range of defect coupling strengths 5\,kHz -- 50\,MHz reported in the literature ~\cite{Martinis2005,Burnett2019,Lisenfeld2019,Abdurakhimov2020}.

The model system Hamiltonian is given in the rotating frame by
\begin{equation}
\hat{H}_S = \hat{H}_0 + \hat{H}_\text{d}(t),
\end{equation}
where the time-independent term is described by
\begin{equation}
\hat{H}_0 = \frac{1}{2} \hbar \Delta \hat{\sigma}_z + i \hbar g \left(\hat{\tau}_- \hat{\sigma}_- - \hat{\tau}_+ \hat{\sigma}_+ \right),
\end{equation}
and the time-dependent drive term is given by
\begin{equation}
\hat{H}_d(t) = \frac{1}{2} \hbar \Omega(t) \hat{O}.
\end{equation}
Here, $\hat{\sigma}_z$, $\hat{\sigma}_\pm = (\hat{\sigma}_x \pm \hat{\sigma}_y)/2$ are Pauli operators of the TLS defect, and $\hat{\tau}_\pm = (\hat{\tau}_x \pm \hat{\tau}_y)/2$ are qubit Pauli operators in the $\{\ket{0},\ket{1}\}$ basis. The operator $\hat{O}$ determines a particular gate type, and, in this work, two types of gates were simulated: the idle gate ($\hat{O} = \hat{I}$) and the $\hat{Y}$ gate ($\hat{O} = \hat{\tau}_y$). The variable $\Omega(t)$ describes the shape of the drive pulse, and, for simplicity, we assume that the pulse has a Gaussian shape described by the pulse duration $t_\text{g}$, time offset $t_\text{off}$, drive amplitude $\Omega_\text{max}$, and standard deviation $\sigma_\text{G}=t_\text{g}/16$:
\begin{equation}
\Omega(t) = \Omega_\text{max} e^{-\frac{(t-t_\text{off})^2}{2\sigma_\text{G}^2}}, \quad 0 \leq t \leq t_\text{g}.
\label{eq-gaussian}
\end{equation}

We calculate the gate error $p$ using the following equation:
\begin{equation}
p = 1 - F\left(\rho(t_\text{g}), \rho_\text{target}\right),
\end{equation}
where $F(\rho(t_\text{g}), \rho_\text{target})$ is the quantum state fidelity of the forward propagated state and target state described by reduced density matrices of the qubit subsystem $\rho(t_\text{g})$ and $\rho_\text{target}$, respectively. The density matrix of the forward propagated state is calculated by numerically solving a Lindblad master equation for the system Hamiltonian $\hat{H}_S$ for given values of the pulse duration $t_\text{g}$, time offset $t_\text{off}$, drive amplitude $\Omega_\text{max}$, and qubit-defect detuning $\Delta$. The initial system state is $\ket{\psi_0} = \ket{1g}$, and the target state is $\ket{\psi_\text{target}} = \hat{O} \ket{\psi_0}$. In reported simulations, we used the qubit energy-relaxation and dephasing rates  $\Gamma_{1\text{q}} =$\,0.01\,$\mu$s$^{-1}$ and $\Gamma_{2\text{q}} =$\,0.01\,$\mu$s$^{-1}$, respectively, and TLS defect energy-relaxation and dephasing rates $\Gamma_{1\text{TLS}} =$\,1\,$\mu$s$^{-1}$ and $\Gamma_{2\text{TLS}}=$\,1\,$\mu$s$^{-1}$, respectively.

\begin{figure}[h!]
    \centering
    \includegraphics[width=\columnwidth]{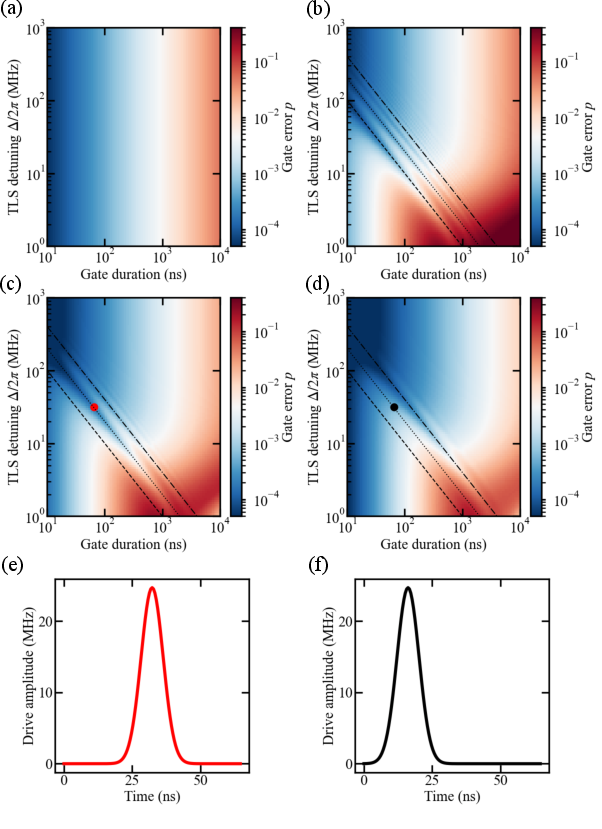}
    \caption{Results of numerical simulations of single-qubit gate errors as a function of the gate duration $t_\text{g}$ and frequency detuning between the qubit and charge-fluctuation TLS defect $\Delta$. (a-b) The error of the idle gate $\hat{I}$ for the qubit-defect coupling strength (a) $g = $\,0 and (b) $g/2\pi =$\,500\,kHz. (c-d) The error of the $\hat{Y}$ gate for time offsets (c) $t_\text{off}=0.5 t_\text{g}$ and (d) $t_\text{off} = 0.25 t_\text{g}$ (in both cases, $g/2\pi =$\,500\,kHz). The red and black dots in~\cref{gate-fidelity}(c,d) represent the same values $t_\text{g}\approx$\,65\,ns and $\Delta/2\pi\approx$\,32\,MHz, and corresponding optimized pulse shapes for given time offsets $t_\text{off}=0.5 t_\text{g}$ and $t_\text{off}=0.25 t_\text{g}$ are plotted in~\cref{gate-fidelity}(e,f), respectively. The dashed, dotted, and dash-dotted lines in~\cref{gate-fidelity}(b-d) correspond to the functions $\Delta = C_\text{g} \times 2\pi/t_\text{g}$ with $C_\text{g}=$\,1, 2, and 4, respectively. }
    \label{gate-fidelity}
\end{figure}

For given values of $t_\text{g}$, $t_\text{off}$, and $\Delta$, we minimize the gate error $p$ by varying the drive amplitude $\Omega_\text{max}$ using the Nelder-Mead optimization method in Python. The initial value of $\Omega_\text{max}$ was equal to $\sqrt{\pi/2}/\sigma_\text{G}$, which corresponded to the $\pi$ rotation of a bare qubit state (as can be shown by integrating~\cref{eq-gaussian}). 

\begin{figure}[h!]
    \centering
    \includegraphics[width=1.0\columnwidth]{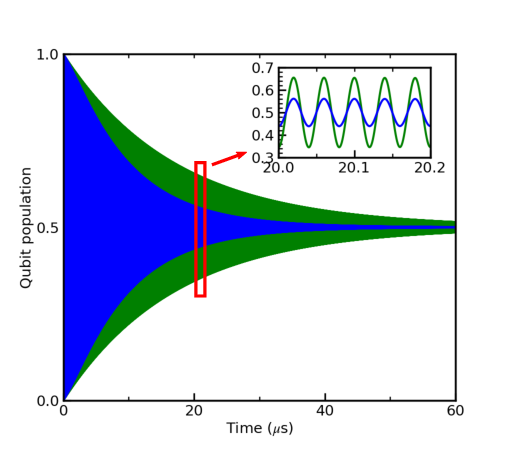}
    \caption{Results of numerical simulations of the qubit evolution under a strong Rabi drive. The population of the qubit state $\ket{0}$ is shown as a function of the Rabi pulse duration. The blue line represents the Rabi decay of a qubit coupled to an off-resonant TLS defect when the pulse amplitude (Rabi frequency) is equal to the qubit-defect detuning ($\Delta_\text{L} = \Omega$). The green curve corresponds to the Rabi decay of a bare qubit ($\Delta_\text{L} \neq \pm \Omega$). Inset: an enlarged view showing Rabi oscillations in the range of pulse duration values close to 20\,$\mu$s.}
    \label{Rabi-simulation}
\end{figure}

Results of the numerical optimization are shown in~\cref{gate-fidelity}. In the case of the idle gate $\hat{I}$ and $g=0$, the gate error increases with the increase of the gate duration due to the bare qubit relaxation [\cref{gate-fidelity}(a)]. In the case of the idle gate $\hat{I}$ and $g=$\,500\,kHz, one can see an oscillation pattern with the characteristic period $2\pi/\Delta$ [\cref{gate-fidelity}(b)]. This oscillation stems from the always-on non-resonant (dispersive) qubit-defect interaction described in the main text. In addition, the increase of the gate error with the decrease of qubit-defect detuning due to the Purcell effect is observed. In~\cref{gate-fidelity}(c,d), we plot results of the optimization of $\hat{Y}$ gates implemented using microwave pulses with two different pulse time offsets $t_\text{off}=0.5t_\text{g}$ and $t_\text{off}=0.25t_\text{g}$, respectively. Examples of optimized pulse shapes at $t_\text{g}\approx$\,65\,ns and $\Delta/2\pi\approx$\,32\,MHz for given time offsets are shown in~\cref{gate-fidelity}(e,f), respectively. The reason for simulating $\hat{Y}$ gates with different $t_\text{off}$ is to show that the oscillation patterns are different for different time offsets due to the interplay between the drive-induced rotation of the qubit state vector and its fast small-amplitude precession caused by the dispersive qubit-defect coupling. For a given qubit-defect detuning $\Delta$, the gate error can be minimized by setting the gate duration to the value $t_\text{g}^\text{opt} = C_\text{g} \times 2\pi/\Delta$, where the parameter $C_\text{g}$ depends on details of a particular qubit gate implementation and should be determined numerically. For example, $C_\text{g}\approx$\,1, $C_\text{g}\approx$\,2, and $C_\text{g}\approx$\,4 for the gate error simulations shown in~\cref{gate-fidelity}(b-d), respectively.

\subsection{Qubit-defect coupling in the case of a Rabi drive\label{appendix-Rabi}}
Figure~\ref{Rabi-simulation} shows results of numerical simulations of the qubit evolution under a strong Rabi drive. Here, the qubit was initialized in the state $\ket{0} = (\ket{i-}+\ket{i+})/\sqrt{2}$. Other parameters were the same as used in the simulation presented in~\cref{fig1}(c,d). The decay of Rabi oscillations is affected by the interaction between the qubit and off-resonant TLS defect when the Rabi frequency is equal to the absolute value of the qubit-defect detuning.

% \bibliography{tls_references}
%

\end{document}